\documentclass[acmsmall]{acmart}

\setcopyright{none}
\renewcommand\footnotetextcopyrightpermission[1]{} %
\usepackage{subcaption}
\usepackage{caption}
\usepackage{tikz}
\usepackage{amsmath}
\usepackage{multirow,booktabs}
\usepackage{xspace}
\usepackage{paralist}
\usepackage{makecell}
\usepackage{cleveref}
\usepackage{siunitx}
\usepackage{pifont}
\usepackage{enumitem}
\usepackage{wrapfig}
\usepackage{xcolor} %
\usepackage{lipsum} %

\newcommand{\ricky}[1]{\textit{\textcolor{teal}{#1}}}
\newcommand{\echont}{\emph{NT-A}\xspace}
\newcommand{\johnnt}{\emph{NT-B}\xspace}
\newcommand{\ucsdnt}{\emph{NT-C}\xspace}
\providecommand{\eg}{\emph{e.g.,} }
\providecommand{\ie}{\emph{i.e.,} }
\providecommand{\etc}{\emph{etc.}\xspace}      %
\newcommand{\todo}[1]{\textcolor{red}{TODO: \emph{#1}}}
\newcommand{\todocite}{\textcolor{red}{TODOCITE}\xspace}
\newcommand{\todonum}{\textcolor{red}{TODONUM}\xspace}
\newcommand{\tododate}{\textcolor{red}{TODODATE}\xspace}
\newcommand{\sebastian}[1]{\textit{\textcolor{orange}{Note from Sebastian: #1}}}
\newcommand{\oliver}[1]{\textit{\textcolor{purple}{Note from Oliver: #1}}}
\newcommand{\hammas}[1]{\textit{\textcolor{red}{#1}}}
\newcommand{\kc}[1]{\textit{\textcolor{red}{#1}}}
\newcommand{\para}[1]{{\vspace{0in} \bf \noindent #1 }}
\newcommand{\parait}[1]{{\vspace{0in} \em \noindent #1 }}

\newcommand{\cmark}{\ding{51}}
\newcommand{\xmark}{\ding{55}}
\newcommand{\etal}{et al.\xspace}
\newcommand{\feat}[1]{\textbf{\underline{#1}}}
\newcommand{\finding}[1]{\textbf{\textcolor{red}{#1}}}

\newcommand*\fullcirc[1][1ex]{\tikz\fill (0,0) circle (#1);}
\newcommand*\halfcirc[1][1ex]{%
	\begin{tikzpicture}
		\draw[fill] (0,0)-- (90:#1) arc (90:270:#1) -- cycle ;
		\draw (0,0) circle (#1);
\end{tikzpicture}}

\definecolor{lightgray}{gray}{0.9}
\begin{document}

\noindent
\fcolorbox{black}{lightgray}{
    \parbox{\linewidth}{
        \centering
        \textbf{Preprint for a paper accepted at \textit{ACM CoNEXT 2025} for publication in the \textit{Proceedings of the ACM on Networking.}}
    }
}
\vspace{1em} %
	\date{}

    \title{Unveiling IPv6 Scanning Dynamics: A Longitudinal Study Using Large Scale Proactive and Passive IPv6 Telescopes}

	\author{Hammas Bin Tanveer}
	\affiliation{%
	  \institution{University of Iowa/{CAIDA}}
	  \city{Iowa City}
	  \country{United States}}

	\author{Echo Chan}
	\affiliation{%
	  \institution{Akamai Technologies/Hong Kong Polytechnic University}
	  \city{Cambridge}
	  \country{United States}}

	\author{Ricky K. P. Mok}
	\affiliation{%
	  \institution{{CAIDA}/UC San Diego}
	  \city{La Jolla}
	  \country{United States}}

	\author{Sebastian Kappes}
	\affiliation{%
	  \institution{Max Planck Institute for Informatics}
	  \city{Saarbrücken}
	  \country{Germany}}

	\author{Philipp Richter}
	\affiliation{%
	  \institution{Akamai Technologies}
	  \city{Cambridge}
	  \country{United States}}

	\author{Oliver Gasser}
	\affiliation{%
	  \institution{IPinfo}
	  \city{Seattle}
	  \country{United States}}

	\author{John Ronan}
	\affiliation{%
	  \institution{Walton Institute, South East Technological University}
	  \city{Waterford}
	  \country{Ireland}}

	\author{Arthur Berger}
	\affiliation{%
	  \institution{Akamai Technologies/MIT}
	  \city{Cambridge}
	  \country{United States}}

	\author{kc Claffy}
	\affiliation{%
	  \institution{CAIDA}
	  \city{La Jolla}
	  \country{United States}}

	\begin{abstract}
We introduce new tools and vantage points to develop and integrate proactive techniques to attract IPv6 scan traffic, thus enabling its analysis. By deploying the largest-ever IPv6 proactive telescope in a production ISP network, we collected over 600M packets of unsolicited traffic from 1.9k Autonomous Systems in 10 months. We characterized the sources of unsolicited traffic, evaluated the effectiveness of five major features across the network stack, and inferred scanners' sources of target addresses and their strategies.

 	\end{abstract}
	\maketitle
\section{Introduction}\label{sec:newintro}
Internet scanning is a vital tool for researchers and malicious actors alike. 
Researchers use scanning to characterize network dynamics \cite{durumeric2013zmap,bano2018scanning,markowsky2015scanning} while malicious actors scan 
to understand a network's attack surface. Capturing and analyzing scanning traffic allows network operators and researchers to study scanner behavior and intent, \eg exploitation of specific vulnerabilities, and consequently build effective defenses against potential malicious traffic. 

The advent of IPv6 has increased the complexity of the Internet scanning ecosystem. Brute-force scanning of the entire IPv6 address space is practically impossible due to the vastness of  address space. Thus, IPv6 scanners must intelligently select targets to increase the likelihood of discovering active addresses.  This more sophisticated approach, unlike the brute force approach to IPv4 scanning, renders it harder to detect IPv6
scanning activity.  In IPv4, scanning detection leverages \emph{darknets}: regions of address space that are inactive, \emph{i.e.}, they neither generate traffic nor host services. Although such dark regions of IPv6 address space are plentiful, they are not as effective for detecting scans,
as scanners cannot afford to exhaustively probe such vast inactive address spaces.

This reality leaves us with limited visibility into potentially
malicious IPv6 scanning activities, hindering our ability
to develop methods to secure and protect IPv6 networks, during a time
of continuous growth in both IPv6 usage~\cite{google_v6}
and attempts to exploit it~\cite{CrowdSec32:online,IPv6thre25:online}.
The potential threat posed by IPv6 scanning has also been recognized within the IETF, which published Internet drafts on operational considerations for 
IPv6 blocklisting \cite{draft-ietf-opsec-ipv6-addressing} and a method
for operators to specify end-site prefix lengths of their networks 
to facilitate sensible IPv6 blocklist 
entries \cite{draft-ietf-opsawg-prefix-lengths}.

\begin{figure}[h!]
	\centering
	\includegraphics[width=0.75\linewidth]{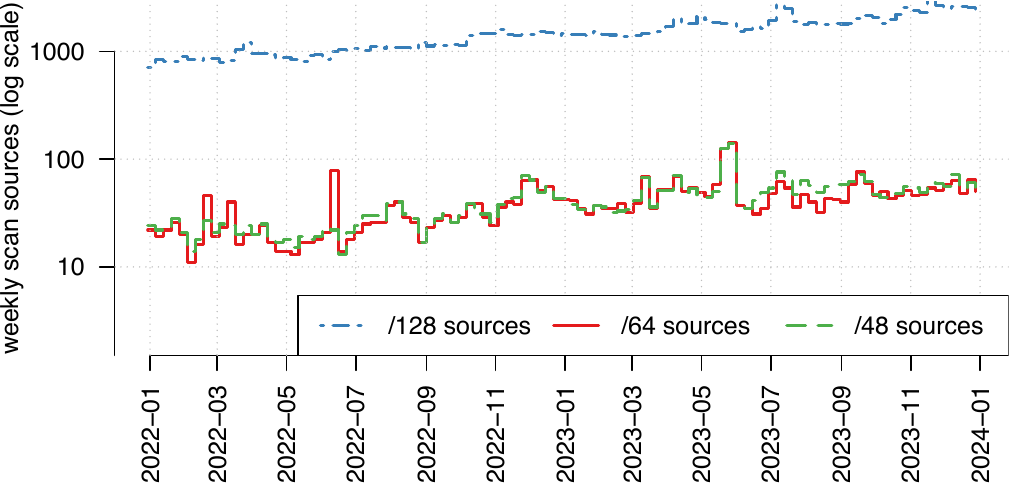}
	\caption{Steady increase in IPv6 scan sources per week hitting the CDN. Number of IPv6 addresses (/128s) more than doubled over the two-year window.  When aggregated into /64, and /48 subnets, the weekly rate tripled, from $\approx$20 to $\approx$50-70.
	}
	\label{fig:source_prefixes_over_time}
\end{figure}
\begin{figure}[h!]
	\centering
	\includegraphics[width=0.75\linewidth]{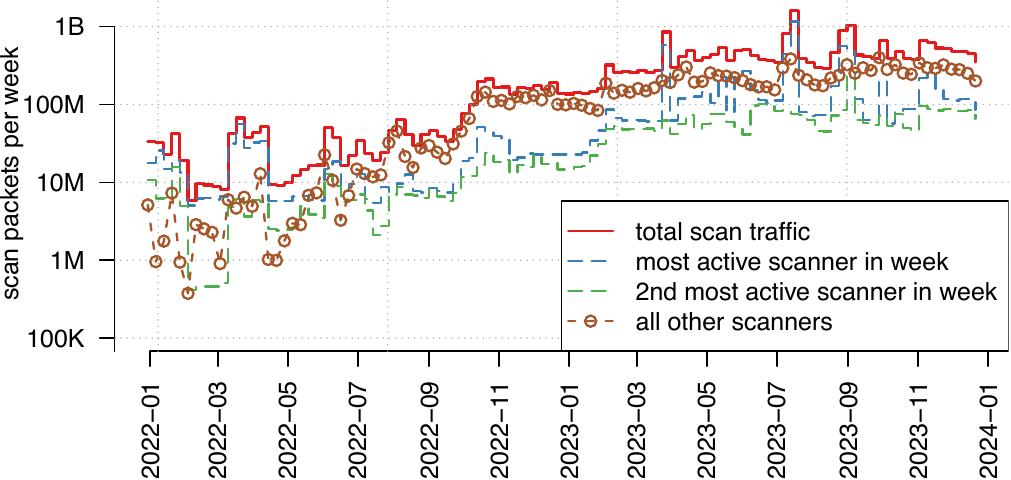}
	\caption{Weekly scan packets (/64 aggregation),
		grew 100X, from 10-60M to 1B.
		In early 2022, scan traffic was often dominated by the
		most active source(s) (dashed lines), by late 2023
		scanning traffic comes from a broad range of sources.}
	\label{fig:scanpkts}
\end{figure}
To validate our assumption that IPv6 scanning traffic is a growing
threat, we collaborated with a major Content Delivery Network (CDN) to capture unsolicited IPv6 packets
reaching a subset of the CDN's servers (230,000 machines in over 700 Autonomous Systems (ASes))
from January 1, 2022 to January 1, 2024.
Figures~\ref{fig:source_prefixes_over_time} and \ref{fig:scanpkts}
show the steady growth in weekly counts of scanning source IPs, 
scanning subnets, and scanning packet volume reaching the CDN over 
this two-year period.\footnote{We define a 
scan as a source hitting at least 100 IPv6 addresses
of the CDN, with a maximum packet inter-arrival time of
3,600 seconds, which allows comparability with earlier
work that used this timeout for IPv6 scan 
detection~\cite{richter2022illuminating}.  
This earlier work analyzed the sensitivity of the 
timeout (3600s, 1800s, 900s) and found only marginal differences,
with scan detection rates declining by single-digit 
percentages under shorter thresholds.
Given this trade-off, and for a lower bound of 100 probed targets 
to declare a scan, a timeout of 1 hour was a reasonable choice for our dataset.
The main point of the figures is not the number of scans, 
but rather its significant increase over time.  
The prefix aggregation accounts for scanners that choose random
source addresses within larger prefixes to evade scan
detection~\cite{richter2022illuminating}.}
(Figure~\ref{fig:source_ases_over_time}
in \Cref{app:cdn} shows a similar growth in number of ASes sourcing
scanning traffic).  
Not only has scanning traffic volume increased two orders of magnitude but it is also more broadly distributed, no longer dominated by one or two scanning sources as it was in early 2022 (see \Cref{app:cdn} for more details).

This growing threat motivates our goal in this study: 
 to develop novel and reproducible techniques to 
attract, capture, and analyze IPv6 scan traffic.
This paper describes our three major contributions:

\noindent \textbf{1. New methods and tools to capture IPv6 scan traffic:} We developed and implemented novel methods to build \emph{proactive telescopes} to attract unsolicited IPv6 scanning traffic (\S\ref{sec:capture}). Our approach not only reacts to incoming traffic, but also emits network signals to attract IPv6 scanners. We designed four proactive attraction features: announcing Border Gateway Protocol (BGP) prefixes, registering domain names, issuing Transport Layer Security (TLS) certificates, inserting addresses in IPv6 hitlists. Also, we deployed reactive features--honeypots of different interaction levels to attract IPv6 scanners.\emph{ To the best of our knowledge, this work integrates the most comprehensive set of features among the prior work we surveyed (\S\ref{sec:related}).}

In collaboration with a regional Internet Service Provider (ISP), we deployed different combinations of features within subnets of their /32 IPv6 address space, namely \emph{honeyprefixes} 
(\S\ref{sec:expr:protelescope}), and compared the results to two passive telescope networks. 
Our experimental design and deployment strategy allowed us to systematically examine the effect size of individual components in attracting scan activities.
We will release the source code of our tools to support reproducibility.

\noindent \textbf{2. Characterization of modern IPv6 scanning:} 
Our proactive techniques induced an increase of three orders of magnitude
in unsolicited IPv6 traffic to our previously un-probed honeyprefixes.
We stratified deployment of techniques across different subnets to
reveal the magnitude and type of IPv6 scanning each technique attracts.
Over 90\% of observed scanning traffic used ICMP ping, even though the telescopes
were responsive to TCP/UDP traffic (\S\ref{sec:results}).  We
also found that scanning traffic/strategies differed vastly by
source ASN; \eg \textit{Internet Scanner} ASNs mostly sent TCP/UDP
packets whereas most scanning traffic from \textit{Hosting/Cloud
providers} was ICMP.

\noindent \textbf{3. Implications to IPv6 security:} Our findings can help to evaluate and improve the efficacy of network security tools to avoid collateral damage. For example, we found scanners used an entire /30 to send scanning traffic, compared to a /96 for some cloud providers. Awareness of address allocation block size is more critical to safely deploying IPv6 blocklists than in IPv4.

\section{Background}
\label{sec:bg}
We provide an overview of the use of network telescopes in capturing unsolicited network packets from darknets, IPv6 Internet scanning techniques, and complexities introduced by IPv6 address allocation and assignment practices to network security tools/techniques.

\subsection{Internet Darknets}
\label{sec:bg_darknets}
Darknets are regions of unused IP address space that do not emit any
network traffic; thus, packets in-bound toward the darknet
are typically part of unsolicited scanning of the address space.
Darknets are used to detect and analyze Internet-wide IPv4 scanning
activities, which became pervasive after the release of 
tools capable of scanning the entire IPv4 address space in minutes \cite{zmapdotio}. 
However, the vast size of IPv6 address space makes such brute-force 
scanning techniques infeasible \cite{Beverly:2018:IBS:3278532.3278559}.
Strategic IPv6 target selection becomes critical 
to maximize the probability of finding responsive addresses. 
Attracting IPv6 scanning activity to darknets requires
simulating network {\em liveness} by generating signals of network activities. 

\subsection{IPv6 Internet Scanning Techniques}
\label{sec:bg_scan_techniques}
An obvious approach to scanning IPv6 relies on curated \emph{hitlists} \cite{gasser2018clusters,zirngibl2022rustyclusters,steger2023targetacquired,Addrminer,rye2023hitlist}.  Researchers have invested in 
methods to improve IPv6 scanning effectiveness 
\cite{williams20246sense,murdock2017target,ullrich2015reconnaissance,6tree,6hit,det,6graph,6gan}, which typically involve two steps:
1) collecting a set of active (seed) IPv6 addresses, 2) generating candidate scanning targets.   
The first step leverages sources of information containing either active IPv6 addresses (\eg \textit{AAAA} records of domains) or hints of address space liveness (\eg BGP announcements). This step yields either exact addresses to scan or a narrowed search space for discovering previously unobserved IPv6 addresses.
The second step uses the data acquired in step 1 to generate (previously unobserved) target addresses with a higher probability of being active. This step 
often involves using machine learning algorithms to find semantic patterns in observed IPv6 addresses and generating candidate addresses with similar patterns (\eg \cite{6gan, williams20246sense}). 

\subsection{Related Work}
\label{sec:related}

Early attempts to capture IPv6 scanning traffic using darknets did not achieve significant visibility \cite{ford2006initial, huston2010background, czyz2013understanding, liu2021darknet, ronan2023revisiting}, motivating more creative approaches.  Fukuda \etal \cite{Fukuda:2018:KID:3278532.3278553} used \emph{DNS backscatter} to identify widespread scanning activity.  
Richter \etal \cite{richter2022illuminating} passively collected unsolicited network packets at a large-scale commercial CDN, uncovering thousands of weekly scan events originating from dozens of different ASes. 

Other methods for capturing IPv6 scanning traffic have \emph{simulated network activity} in darknets to attract scanners. Tanveer \etal \cite{tanveer2023glowing} launched services from an unused /56 IPv6 prefix to indicate a subnet's ``liveness'', resulting in an increase of 
several hundreds of scanning packets per day. Zhao \etal \cite{Zhao2024v6darknet} uncovered how scanners utilized DNS-based methods to scan IPv6, \eg enumerating IPv6 addresses by walking the ip6.arpa zone or finding AAAA records using IPv4 PTR records. They found that most IPv6 scanners they attracted had
discovered target IPv6 addresses using IPv4 PTR records.
Both Scheitle \etal \cite{scheitle2018rise} and Pletinckx \etal \cite{pletinckx2023ctlogs} showed that exposing a domain name in Certificate Transparency Logs likewise induced traffic from scanners. Egloff \etal \cite{ehksw-dmvis-25} uncovered how IPv6 scanners react to BGP prefix announcements. 

We build on this previous work to create a more systematic, comprehensive, scalable, and reproducible approach to soliciting IPv6 scanning traffic.   We use a larger (/32) network than  previous studies and subdivide it into smaller subnets, each employing different combinations of techniques, to isolate the effects of individual variables in our experiment. 

\section{Methodology}
\label{sec:capture}

Both prior work and our experiments (\S\ref{sec:expr:telescope}) showed that IPv6 prefixes that emit little to no traffic receive substantially less unsolicited network traffic
compared to active prefixes
\cite{tanveer2023glowing,Zhao2024v6darknet}. 
Our approach leverages both passive as well as \textit{proactive} methods to attract and capture unsolicited IPv6 traffic. The passive approach deploys darknet-based network telescopes to capture unsolicited traffic. We design and implement a novel \emph{proactive telescope}, which not only reacts to incoming traffic but also stimulates Internet scanners that use public data sources to find probing targets.

\subsection{Darknet-based Network Telescopes}
We deploy darknet-based network telescopes in two ways to establish the baseline for unsolicited traffic. Our first approach monitors the ingress traffic to a dedicated network prefix over time, similar to Czyz et al. \cite{czyz2013understanding,ronan2023revisiting}. 
The second approach captures traffic destined to unused portions of a live network. 
We leverage the internal routing table of the network's border router
to forward ingress traffic destined to these unused subnets.
The size of the telescope is dynamic, depending on 
the current subnet assignment within the network.

\subsection{Proactive Network Telescope: Exposing IPv6 Addresses to Public Data Sets}
\label{sec:method:proactive}

We develop a \textit{proactive network telescope} by \textbf{1)} increasing the visibility of our network telescope's address space to potential scanning actors by exposing the telescope's IPv6 range in public datasets and \textbf{2)} emulating live network services by reacting to incoming traffic.
We advertise prefixes and addresses used by the telescope to the Internet using multiple network protocols, \eg BGP, DNS, and TLS, resulting in their inclusion
in publicly available datasets, \eg RIPE Routing Information Service (RIS) \cite{riperis}, DNS zone files \cite{zonfileaccess} \etc, %
thereby signaling evidence of network activity. 

Our work improves the deployment of methods proposed in prior research \cite{tanveer2023glowing, Zhao2024v6darknet} in two ways.
First, our proactive telescope utilizes a significantly larger address space --- a /32 compared to a /56 in previous studies --- which enables us to both capture more unsolicited traffic, 
and to examine scanning triggers that were previously inaccessible. Second, we conduct a more comprehensive set of active experiments to attract scanning activity \eg advertising our prefixes via TLS certificates, compared to previous works that primarily relied on like Certificate Transparency (CT) logs \cite{scheitle2018rise, pletinckx2023ctlogs, kondracki-usenix22}. This approach allows us to gain deeper insights into how scanners integrate diverse datasets into their scanning strategies.

Specifically, we conduct the following active experiments in our proactive network telescope.

\noindent \textbf{BGP announcements.}
We randomly select /48 prefixes from the upper half\footnote{The ISP requested we use the upper half, and prefix location does not affect generalizability of our study. (\Cref{fig:prefix_announce_map}).} of the ISP's /32 and announce them via BGP to create \textit{honeyprefixes}. 
Then, we verify that the selected prefixes received little to no traffic in the prior month.
We choose the /48 as it is the longest prefix that reliably propagates globally \cite{sediqi2022hyper}, allowing us to maximize utilization of the /32 address space. 

The use of BGP announcements gave us two advantages over previous works \cite{tanveer2023glowing,Zhao2024v6darknet}: 1) it signals to scanners that these regions are active (scanners can monitor such announcements using BGP route collectors like RouteViews \cite{routeviews}; and (2) it provides a clear boundary of address space and significantly reduces the search space compared to the covering /32 prefix, increasing the likelihood of scanners discovering active addresses. 
We conducted all subsequent active experiments within these BGP-announced /48 \textit{honeyprefixes}.

We announce the first subnet in some selected prefixes with a longer prefix length (/49-/64) to examine if scanners can discover prefixes with low visibility in routing tables.
We record the time when the routes propagate to public route collectors as the start point for our experiments.

\noindent \textbf{Domain names.}
We register new domain names in multiple top-level domains (TLDs) and created AAAA records pointing the root domain (\ie eTLD+1 domain) to random IPs in our honeyprefixes, exposing them to TLD and domain monitoring tools.
Scanners that monitor TLD zone files could resolve the domains names to IPs and subsequently probe our honeyprefixes.

\noindent \textbf{Subdomain names.}
\label{fig:capture:subdomain}
Network operators commonly use subdomain names (\ie eTLD+2 domains) for various services, such as \text{www}, \texttt{mail}, and \texttt{ns}, by convention. We select a total of 374 names listed on at least three of four popular subdomain name lists \cite{commonspeakwordlist,dnsscan,dnspop,dnslist}. We deploy \texttt{AAAA} records to map each subdomain name to a randomly assigned IPv6 address within a honeyprefix.
\noindent \textbf{TLS certificates.}
As most websites adopt HTTPS, we issue TLS certificates for the root and 50 randomly chosen subdomain names to mimic the presence of web services. CT logs \cite{ctlog} reveal  the existence of these subdomains without access to their DNS zone files, allowing scanners to discover subdomains listed in the certificates.

\noindent \textbf{IPv6 hitlist.}
IPv6 hitlists (\S\ref{sec:related}) periodically compile lists of responsive targets using public datasets and active measurement results. 
Our honeyprefix addresses were thus discovered by and then appeared in such
hitlists. 
We also collaborated with a major hitlist provider to manually add %
random addresses from two honeyprefixes that would not otherwise be
discovered by its process. This allowed us to identify scanners that rely on hitlists as their source of probing targets.
\subsection{IPv6-Native Interaction Honeypots}
\label{sec:method:twinklenet}

Our proactive telescope engages with incoming IPv6 probes using both low and high interaction honeypots, as scanners may elicit scanning behavior that 
darknet telescopes cannot observe \cite{hiesgen22spoki}.
We seek to explore the efficacy of such interactions for IPv6 telescopes.

\textbf{Twinklenet.}
We designed a lightweight low-interaction multi-protocol honeypot called \emph{Twinklenet} to respond to unsolicited incoming traffic sent to dedicated subnets and/or IPs.
Existing open-source honeypots (\eg AmpPot \cite{Kramer15ampot}, T-pot \cite{tpot}, and Spoki \cite{hiesgen22spoki}) support neither IPv6 nor multi-protocol IP aliasing (\ie handling packets to multiple destination addresses with a single network interface). 
An alternative approach to enable IP aliasing is to use Network Address Translation (NAT) which does not easily scale in IPv6.

Twinklenet supports IP aliasing for both IPv4 and IPv6 address spaces and responds to four popular protocols (see \Cref{app:twinklenet}). 
A single instance of Twinklenet can handle incoming packets toward multiple non-contiguous subnets and IPs. 
In addition to responding to ICMP pings,
Twinklenet can bind to any TCP port of any honeyprefix's subnets and IPs  to accept incoming connections. After completing the TCP handshake, and capturing the first data packets sent by the scanner, Twinklenet gracefully closes the connection. 
For UDP, crafting responses requires parsing protocol-specific queries.
Twinklenet supports two popular UDP-based protocols: DNS and NTP. Instead of implementing full services, which attackers may exploit for abuse, Twinklenet replies with  error messages to indicate responsiveness to the sender. We plan to make Twinklenet available
to the research community upon the acceptance of this paper.

\textbf{High-Interaction Honeypots.}
To examine whether scanners behave differently with full-stack systems, we integrated a high-interaction honeypot into our proactive telescope using T-Pot~\cite{tpot}, a container-based framework emulating various services. Since each T-Pot instance can only bind to a single IPv4 address, we designed a two-stage setup to enable IPv6 and address aliasing across an entire honeyprefix (see \Cref{app:tpot}, \Cref{fig:tpotinfra}).

Traffic to the honeyprefix is redirected via an access router, which maps all packets to the prefix's first address (\texttt{::1}) and source port using DNAT. The traffic is then forwarded to a reverse proxy that performs static 6-to-4 translation to T-Pot's IPv4 address and routes it to the appropriate container based on protocol and port.
We log DNAT mappings (timestamp, original IPv6 destination, source port) to recover original destinations in T-Pot logs. The access router also mirrors traffic to a packet capturer, storing it in PCAP format.

\subsection{Quantifying effects of controlled experiments}
\label{sec:capture:quantify}

We then measure the impact of our controlled experiments in attracting IPv6 scanning traffic.

\textbf{Treatment vs. Control Subnets.} 
We divide our /32 address space into two groups: \textit{treatment} subnets (\ie \emph{honeyprefixes}), and the remaining /48 prefixes (within the covering /32) as control subnets.
As the BGP announcement is the initial feature to all honeyprefixes,  a control subnet becomes a treatment subnet at the time its BGP announcement is observed by public route collectors.

\textbf{Quantification Method.} We use Bayesian structural time-series models (\textit{BSTM})~\cite{brodersen2015inferring} to quantify the effects of our experiments. We use BSTM to construct the counterfactual ---  what scanning activity would have looked like in the \textit{Treatment} subnet, had the intervention (controlled experiment) not been applied -- for each \textit{honeyprefix}. Each counterfactual takes in two inputs; (1) the pre-treatment scanning activity in the treatment subnet, and (2) the scanning activity in the control subnet that received the most scanner attention during the experiment. This approach
ensures that we calculate the lower bound of the effect of our experiments.

Using BSTM over traditional Difference-in-Difference method gives us 3 major advantages; First, BSTM enables us to capture complex scanning behavior \eg sudden bursts of activity, yielding more accurate time series for the controls (prefixes for which we did not deploy features). Second, it does not expect parallel trends \ie control and treatment to evolve similarly over time. This aspect is important as previous works \cite{tanveer2023glowing} have shown that external factors \eg subnet location in the covering prefix, address structure, \etc. can also influence scanning activity. Third, BSTM can generate dynamic counterfactuals by tweaking the influence of each 
input, which results in robust uncertainty quantification.

\begin{figure}[h]
	\centering
	\includegraphics[width=.8\textwidth]{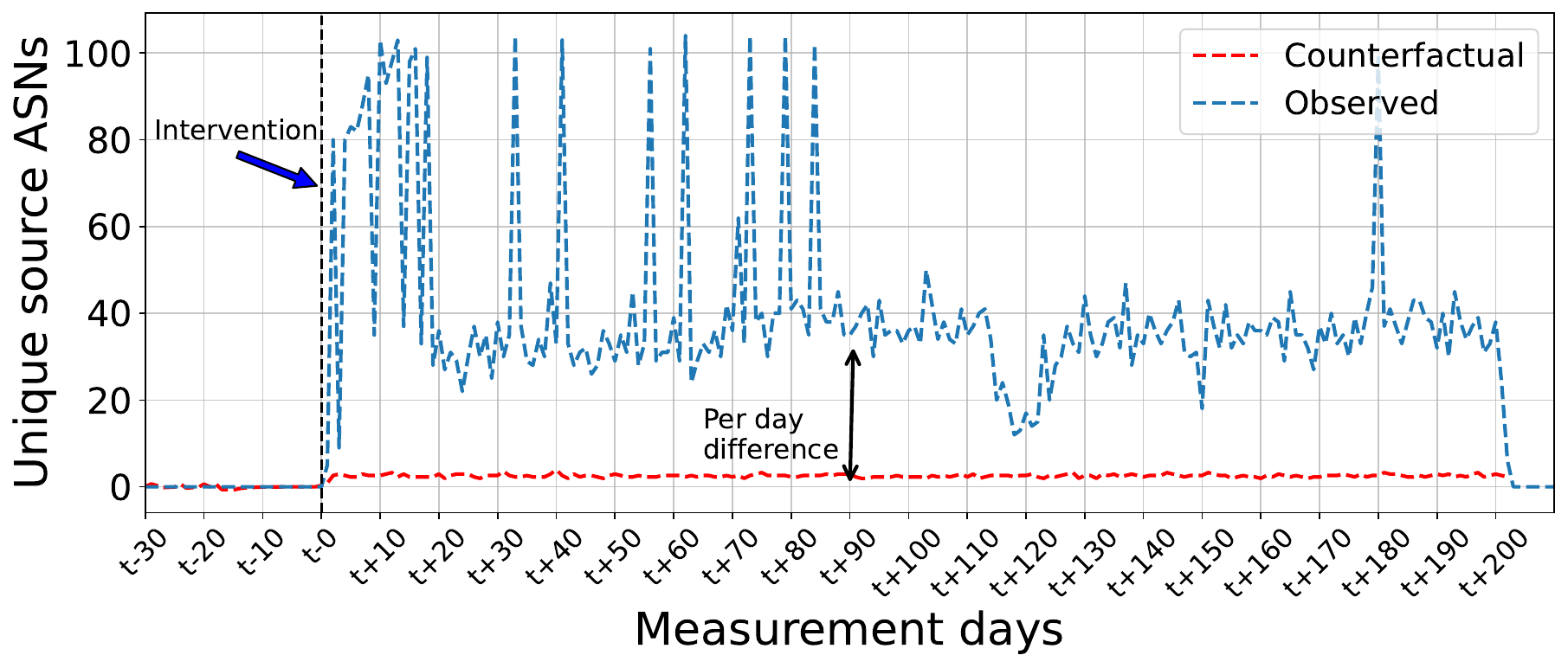}
	\caption{Counterfactual vs. Observed unique ASNs for a honeyprefix}
	\label{fig:bstm}
\end{figure}

\textbf{Calculating Effect Sizes.} We compute the Average Effect Size (AES) as the mean of daily difference between scanning activity in the treatment subnet and counterfactual -- as shown in \Cref{fig:bstm} -- over all days an experiment is active. We focus on two metrics: the AES for traffic volume, denoted as ${\Delta_{\mathcal{H}}^{traffic}}$, and the AES for the number of unique source ASNs, denoted as ${\Delta_{\mathcal{H}}^{ASN}}$. These metrics allow us to quantify both the volume of scanning activity and the diversity of scanners attracted by our experiments. To establish the statistical significance of change in scanners/activity, we compute 95\% confidence intervals by resampling scanning activity pre and post intervention.

\label{app:bstm}

\section{Telescopes and Datasets}
\label{sec:expr}

We next describe our data capture method and datasets gathered from our three
vantage points.

\subsection{Passive Network Telescopes}
\label{sec:expr:telescope}

We deployed three geographically and topologically diverse IPv6 network telescopes (\Cref{tab:telescope_summary})---one in a transit ISP and two in academic networks---to capture unsolicited traffic to the unused addresses. Over our three overlapping collection windows (10 months, 16 months, and 7 months), 
we captured over a billion unsolicited packets 
from $2000$ unique ASs targeting more than 150M unique destinations in all of our telescopes combined.

\begin{table}[h!]
\caption{Overview of scanning traffic/sources captured and destinations targeted in \emph{NT-A}, \emph{NT-B} and \emph{NT-C}.}
\label{tab:telescope_summary}
\resizebox{\textwidth}{!}{%
\begin{tabular}{c| c c c c c cccc ccc}
\toprule
\multirow{2}{*}{\textbf{Telescope}} & \multirow{2}{*}{\textbf{\begin{tabular}[c]{@{}c@{}}Address \\ space\end{tabular}}} & \multirow{2}{*}{\textbf{Location}}                       & \multirow{2}{*}{\textbf{\begin{tabular}[c]{@{}c@{}}Network \\ type\end{tabular}}} & \multirow{2}{*}{\textbf{\begin{tabular}[c]{@{}c@{}}Measurement\\ start-end date\end{tabular}}} & \multirow{2}{*}{\textbf{\begin{tabular}[c]{@{}c@{}}Packets \\ received\end{tabular}}} & \multicolumn{4}{c}{\textbf{Unique traffic sources}}                                                                      & \multicolumn{3}{c}{\textbf{\begin{tabular}[c]{@{}c@{}}Unique dest. \\ targeted\end{tabular}}}      \\ \cmidrule(lr){7-10} \cmidrule(lr){11-13}
                                    &                                                                                    &                                                          &                                                                                   &                                                                                            &                                                                                                   & \multicolumn{1}{c}{\textbf{/128}} & \multicolumn{1}{c}{\textbf{/64}} & \multicolumn{1}{c}{\textbf{/48}} & \textbf{ASes} & \multicolumn{1}{c}{\textbf{/128}} & \multicolumn{1}{l}{\textbf{/64}} & \multicolumn{1}{l}{\textbf{/48}} \\ \midrule
\textit{\textbf{NT-A}}              & /32                                                                                & \begin{tabular}[c]{@{}c@{}}Southern \\ Asia\end{tabular} & \begin{tabular}[c]{@{}c@{}}Transit \\ ISP\end{tabular}                            & 07/23 - 04/24                                                                                     & 654M                                                                                              & \multicolumn{1}{c}{259k}          & \multicolumn{1}{c}{190k}         & \multicolumn{1}{c}{138k}         & 1.9k         & \multicolumn{1}{c}{134M}          & \multicolumn{1}{c}{3.1M}         & 61.5k                             \\
\textit{\textbf{NT-B}}              & /48                                                                                & Ireland                                                  & Research                                                                          & 01/23 - 04/24                                                                                    & 300k                                                                                              & \multicolumn{1}{c}{1.9k}          & \multicolumn{1}{c}{367}          & \multicolumn{1}{c}{354}          & 60           & \multicolumn{1}{c}{100k}          & \multicolumn{1}{c}{65.5k}        & 1                                 \\
\textit{\textbf{NT-C}}              & /32                                                                                & \begin{tabular}[c]{@{}c@{}}United \\ States\end{tabular} & Academic                                                                          & 10/23 - 04/24                                                                                     & 250M                                                                                              & \multicolumn{1}{c}{57k}           & \multicolumn{1}{c}{26k}          & \multicolumn{1}{c}{24k}          & 507          & \multicolumn{1}{c}{21M}           & \multicolumn{1}{c}{14.9M}        & 48.8k                             \\
\bottomrule
\end{tabular}%
}
\end{table}

\textbf{\echont} is hosted in an ISP network in Southern Asia with low IPv6 address space 
utilization. 
The ISP's equipment and its customers use the initial five /48s of the APNIC-assigned /32 block. 
We also collaborate with this ISP to deploy our proactive network telescope (\S\ref{sec:expr:protelescope}).

\textbf{\johnnt} is an Irish IPv6 research network telescope \cite{ronan2023revisitinganon}  monitoring incoming traffic to an unused /48 network since June 2022. 

\textbf{\ucsdnt} is deployed at a U.S. academic network with a /32 assignment from ARIN. Similar to \echont, it captures all the traffic sent to any unassigned subnets within the address space.  
The university has assigned the top half (a /33) of the /32 block to equipment and departments on campus.

\subsection{Proactive Network Telescope Deployment}
\label{sec:expr:protelescope}

\begin{figure}[h!]
	\centering
	\includegraphics[width=.78\textwidth, trim=0.1cm 0.1cm 0.1cm 0.1cm, clip]{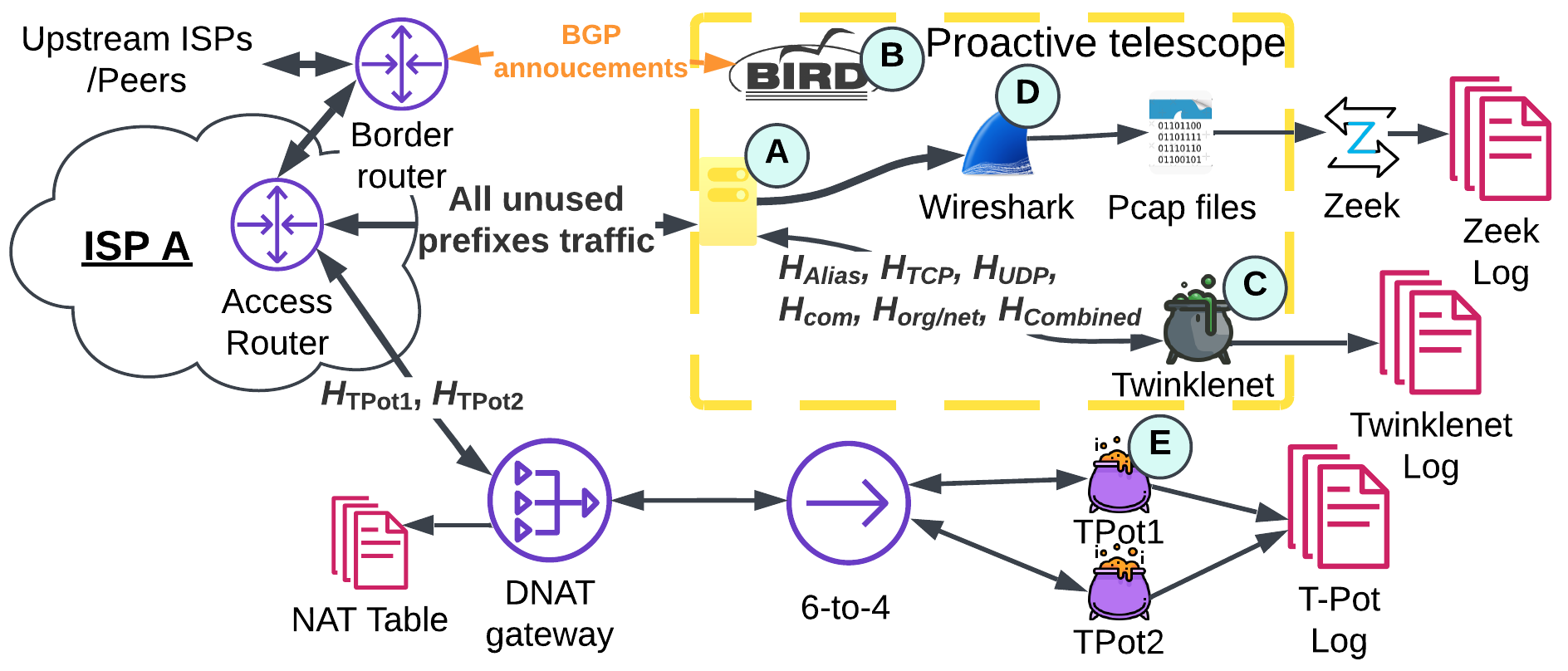}
	\caption{Overview of our IPv6 proactive telescope setup in ISP A. }
	\label{fig:setupoverview}

\end{figure}

We implemented our proactive network telescope in the address space and infrastructure of \echont (\Cref{fig:setupoverview}) as described in \S\ref{sec:method:proactive} and \S\ref{sec:method:twinklenet}. The access router forwarded ingress traffic destined for unused prefixes in ISP A to server \textcircled{A}, which performed three main functions: 1) Run the BIRD \textcircled{B} Internet routing daemon to announce honeyprefixes in ISP A's address space, 2) execute Twinklenet \textcircled{C} to respond to incoming traffic based on the experiment's configuration (Table \ref{tab:honeyfixconfig}), and 
3) capture incoming traffic \textcircled{D} toward unused address space and honeyprefixes.
ISP A's operators register the honeyprefixes on APNIC's Resource Public Key Infrastructure (RPKI) portal, so that the upstreams and peers accepted and propagated the new routes. 
To accommodate the high computational and memory demands, we deployed dedicated machines \textcircled{E} to host two T-Pot instances. They reacted to traffic targeting $\mathcal{H}_{TPot1}$ and $\mathcal{H}_{TPot2}$. %
In order to map the IPs in T-Pot logs, we collected the NAT table from the DNAT gateway containing the mapping records with timestamps.

\subsection{Honeyprefix features and configurations}
\label{sec:expr:protelescope:feat}

We describe the implementation of seven features that we experimentally deployed, in multiple combinations, to infer which data sources scanners use to create their target lists. 
Specifically, we deployed 27 honeyprefixes with different configurations (Table \ref{tab:honeyfixconfig}) to investigate how Internet scanners discover live hosts and react to different network behaviors. 

\subsubsection{Domain and subdomain names} 
\label{sec:expr:protelescope:names}
We purchased a total of 9 domain names (6 \texttt{.com}, 2 \texttt{.net}, and 1 \texttt{.org} domain) from GoDaddy \cite{godaddy}. Shortly after registration, we used the registrar-provided DNS server to set up the root (\ie \texttt{@}) AAAA record for the corresponding DNS zones. Additionally, we deployed AAAA records of over 300 common subdomain names (\S\ref{fig:capture:subdomain}) in 4 of the 9  domain names. All records pointed to a randomly selected IP address in the associated honeyprefix. We used \feat{D} and \feat{S} to denote the domain and subdomain name features, respectively.

\subsubsection{TLS certificates} 
\label{sec:expr:protelescope:tls}
As both low and high interaction honeypots (\S\ref{sec:method:twinklenet}) do not fully emulate an actual web server, we could not use the \texttt{HTTP-01} challenge \cite{letsencryptchallenge}, which requires hosting a special file on the web server to validate our control over the domain names. Instead, we issued TLS certificates using the \texttt{DNS-01} challenge \cite{letsencryptchallenge} with our customized \texttt{certbot} plugin supporting our domain registrar's APIs, enabling automatic insertion of \texttt{TXT} DNS records required by the challenge.
We issued TLS certificates using Let's Encrypt for all the root domain names and only 50 subdomain names (due to Let's Encrypt's weekly certificate limit \cite{letencryptlimit}). We denote the TLS certificate feature for the root name and subdomains with \feat{d} and \feat{s}, respectively.

\subsubsection{IP aliasing}
\label{sec:expr:protelescope:aliasing}
$\mathcal{H}_{Alias}$ and the two honeypots ($\mathcal{H}_{TPot1}$ and $\mathcal{H}_{TPot2}$) 
implemented IPv6 aliasing using Twinklenet and the NAT gateway, respectively. All addresses in these prefixes responded to incoming ICMP echo requests. 

\subsubsection{ICMP responsiveness} 
\label{sec:expr:protelescope:icmp}
We configured Twinklenet to respond to ICMP Echo requests for the first address (`::1') and two randomly selected addresses in non-aliased honeyprefixes ($\mathcal{H}_{RDNS}, \mathcal{H}_{TCP}$, and $\mathcal{H}_{UDP}$). One random address in $\mathcal{H}_{Combined}$ is also responsive to ICMP. We denote the ICMP responsiveness feature to individual IPs and aliased prefixes with \feat{I}.

\subsubsection{TCP/UDP open ports} 
\label{sec:expr:protelescope:ports}
Our honeypot deployment (Twinklenet and T-Pot) reacted to incoming TCP and UDP traffic to specific IPs in the honeyprefixes. We denote the TCP/UDP reactive features as  \feat{T} and \feat{U}, respectively.
We used Twinklenet to simulate popular services over TCP (web, and remote control) and UDP (DNS and NTP) in one randomly selected address in the honeyprefix, respectively. We also enabled web-related ports (TCP 80, 443, 8080, 8443) on the IPs pointed to by AAAA records of domain/subdomain names in $\mathcal{H}_{Com}$, $\mathcal{H}_{Org/net}$, and $\mathcal{H}_{Combined}$. 

We integrated multiple features in $\mathcal{H}_{Combined}$. The first address of $\mathcal{H}_{Combined}$ responded to all ICMP and TCP/UDP common ports. We selected four random IPs to respond to ports 
related to web, remote control-related, DNS, and NTP services, respectively. 
$\mathcal{H}_{TPot1}$ and $\mathcal{H}_{TPot2}$ responded on TCP/UDP ports corresponding to some of the popularly targeted protocols, \eg SSH, Telnet, DNS, and SMTP (see \Cref{app:tpot}, \Cref{table:docker_ports} for the full lists).

\begin{table}[h!]
	\caption{Configuration of honeyprefixes.
	}
	\small
	\label{tab:honeyfixconfig}
	\resizebox{\textwidth}{!}{%
	\begin{tabular}{c|cccccccc}
		\toprule
		Honeyprefixes & \hyperref[subsubsec:control]{BGP} & \hyperref[sec:expr:protelescope:aliasing]{Aliased} & \hyperref[sec:expr:protelescope:icmp]{ICMP} & \hyperref[sec:expr:protelescope:ports]{TCP} & \hyperref[sec:expr:protelescope:ports]{UDP} &  \hyperref[sec:expr:protelescope:names]{Domain} & \hyperref[sec:expr:protelescope:names]{Subdomain} & \hyperref[subsubsec:hitlist]{IPv6 Hitlist} $^\star$ \\ \hline
		Description & \S\ref{subsubsec:control}&\S\ref{sec:expr:protelescope:aliasing}&\S\ref{sec:expr:protelescope:icmp}&\multicolumn{2}{c}{\S\ref{sec:expr:protelescope:ports}}&\multicolumn{2}{c}{\S\ref{sec:expr:protelescope:names} and \S\ref{sec:expr:protelescope:tls}}&\S\ref{subsubsec:hitlist}\\
		\midrule
		$\mathcal{H}_{Alias}$ & 1$\times$/48 &\cmark& \fullcirc& \xmark &\xmark &\xmark &\xmark &Aliased\\
		$\mathcal{H}_{TCP}$ & \xmark$^\ddagger$ &\xmark & \halfcirc & web, remote& \xmark &\xmark &\xmark &\xmark\\
		$\mathcal{H}_{UDP}$ &  1$\times$/48 &\xmark & \halfcirc & \xmark& 53, 123 &\xmark &\xmark &NA, UDP53, ICMP\\
		$\mathcal{H}_{Com}$ &  1$\times$/48 &\xmark & \xmark & web & \xmark  &2$\times$\texttt{.com} &\xmark &NA, TCP80, TCP443\\
		$\mathcal{H}_{Org/net}$ &  1$\times$/48 &\xmark & \xmark & web & \xmark  &1$\times$\texttt{.org}, 1$\times$\texttt{.net} &\cmark (only \texttt{.net}) &NA, TCP80, TCP443\\
		$\mathcal{H}_{Combined}$ &  1$\times$/48 &\xmark & \halfcirc & web, remote & 53, 123  &1$\times$\texttt{.net} &\cmark &NA, TCP80, TCP443\\
		$\mathcal{H}_{TPot1}$ &  1$\times$/48 &\cmark& \fullcirc & \multicolumn{2}{c}{See \Cref{app:tpot}, \Cref{table:docker_ports}}  &2$\times$\texttt{.com} & 1$\times$\texttt{.com} & Aliased, Manual\\
		$\mathcal{H}_{TPot2}$ &  1$\times$/48 &\cmark& \fullcirc & \multicolumn{2}{c}{See \Cref{app:tpot}, \Cref{table:docker_ports}}  & 2$\times$\texttt{.com} & 1$\times$\texttt{.com} & Aliased, Manual\\
		$\mathcal{H}_{Specific}$ & /49-/64 & \xmark &\xmark&\xmark&\xmark&\xmark&\xmark&\xmark\\
		$\mathcal{H}_{BGP}$&3$\times$/48& \xmark &\xmark&\xmark&\xmark&\xmark&\xmark&\xmark\\
		\bottomrule
		\multicolumn{9}{l}{Note: \fullcirc/\halfcirc represent the entire subnet/specific addresses were responsive to ICMP, respectively. }\\
		\multicolumn{9}{l}{~~$\ddagger$: We configured BIRD to announce the prefix, but the announcement failed to reach the Internet due to a technical problem.}\\
		\multicolumn{9}{l}{~~$\star$: Aliased/NA represents the aliased/non-aliased prefixes list, respectively. ICMP, TCP80, TCP443, and UDP53 denote the hitlists that}\\
		\multicolumn{9}{l}{~~~~ reported at least one IP in the subnet as responsive to the corresponding protocol. }\\
		\multicolumn{9}{l}{~~web: 80, 443, 8080, 8443; remote: 22, 23, 2323, 3389.}
	\end{tabular}
	}
\end{table}

\subsubsection{IPv6 Hitlist}
\label{subsubsec:hitlist}
The hitlist measurements \cite{gasser2018clusters} discovered some prefixes and IPs in honeyprefixes. The hitlist's aliased/non-aliased prefix list included all three aliased ($\mathcal{H}_{Alias}$, $\mathcal{H}_{TPot1}$, and $\mathcal{H}_{TPot2}$) and five non-aliased (	$\mathcal{H}_{TCP}$, $\mathcal{H}_{UDP}$, $\mathcal{H}_{Com}$, $\mathcal{H}_{Org/net}$, and $\mathcal{H}_{Combined}$ ) honeyprefixes, respectively.
The hitlist also revealed some IPs with open ports on UDP port 53, and TCP port 80 and 443. We collaborated with the hitlist maintainers to manually add two IPs (one at the beginning of the address space, and one random in the honeyprefix) into each hitlist category. In total, we manually insert 40 addresses (20 per honeypot) across 10 hitlist categories. We label this feature as \feat{H}.

\subsubsection{BGP only prefixes}
\label{subsubsec:control}
We announced 19 BGP only honeyprefixes (3 $\mathcal{H}_{BGP}$ and 16 $\mathcal{H}_{Specific}$). None of them responded to incoming traffic. The three $\mathcal{H}_{BGP}$ prefixes were identical and were announced with a prefix length of /48. We announced a set of 16 $\mathcal{H}_{Specific}$ prefixes with lengths ranging from /49 to /64, with one prefix for each length. %
We refer to the remaining unused dark address space announced through ISP A's covering /32 prefix as $\mathcal{H}_{Control}$.

\subsection{Metadata and data processing} We map source IPs to ASs and countries using CAIDA's RouteViews Prefix-to-AS mapping \cite{prefixtoas} and IPinfo's geolocation database \cite{ipinfogeoip}
with datasets collected on the same day as the packet timestamps, ensuring the timeliness of the mapping. 
We use ASdb \cite{Ziv2021} to identify the type of AS and Zeek \cite{zeek} to aggregate packets into flows to facilitate our analysis.

\section{Results}
\label{sec:results}

We present a comparative analysis of the scanning sources attracted by 3 of our telescopes (\S\ref{sec:comparison}). We conduct a deep dive into scanning traffic collected by our proactive telescope \emph{NT-A} from July 2023 to April 2024. 
We characterize scanning sources and properties (\S\ref{sec:results:traffic}), 
describe our method for setting up our controlled experiments and 
quantifying their effects (\S\ref{sec:results:effect}),
and outline the different 
scanning strategies we observe (\S\ref{sec:results:scanstrategy}).

\subsection{Telescope comparison}
\label{sec:comparison}
To evaluate the efficacy of our proactive telescope (\emph{NT-A}) in attracting scanning traffic, we compare the characteristics of the scan sources across all three telescopes.
Overall, \emph{NT-A} accounted for almost 70\% of all unsolicited traffic that we captured, with  98.4\% of that traffic targeting honeyprefixes.
It also attracted scanning traffic from the most diverse set of sources among the three. \emph{NT-C} received most of the remaining $\sim$30\% of the traffic but from a much smaller set of sources. \emph{NT-B} captured only a small fraction of the total traffic, owing to its order of magnitude smaller address space.

We calculated the \textit{Jaccard similarity} to enumerate the overlap of scanning sources between the 3 telescopes. We calculate \textit{Jaccard Similarity} for scan sources at 3 different prefix lengths, /32, /63 and /128, as follow: \emph{JS}($NT_y^{agg}$,$NT_x^{agg}$) = 
$\frac{|Sources_{agg} NT_{y} \cap Sources_{agg} NT_x|}
{|Sources_{agg} NT_y \cup Sources_{agg} NT_x|}$. The average Jaccard similarity across prefix aggregation levels of /32,
/64 and /128, across all telescope combinations, was $\sim$0.1, indicating 
that the sets of sources observed at different telescopes were highly
distinct. The highest \emph{JS} of 0.2 was observed between \emph{NT-A}
and \emph{NT-C} at a /32 aggregation.

Despite the overall dissimilarity, a few overlapping sources accounted for most of the unsolicited traffic, and targeted the largest number of unique destinations within our telescopes. Comparing source IPs (/128) targeting \emph{NT-A} \emph{vs.}~\emph{NT-B}, the overlapping sources generated 4.3\% of unsolicited traffic received by \emph{NT-A}. This fraction increased  to 96.3\% when we aggregated by a longer prefix length \textit{i.e.,} /64. Common sources between \emph{NT-A} and \emph{NT-C} aggregated by /64 generated 97.3\% and 99.2\% of the unsolicited traffic received by \emph{NT-A} and \emph{NT-C}, respectively. 

\emph{NT-A} exclusively captured the most exploratory scanners \ie scanners that were responsible for targeting the most unique destination IPs. For example, overlapping sources (grouped by /64) between \emph{NT-A} and \emph{NT-B} were responsible for most unsolicited traffic (96.8\%) in both telescopes, but these sources targeted only 45.1\% of all unique destination IPs probed in \emph{NT-A}.

Hosting/cloud providers were responsible for the majority ($>50\%$) of unsolicited traffic sent toward 
\emph{NT-A} and \emph{NT-C}, which together received over 99\% of all unsolicited traffic across the three telescopes.
Notably, \emph{Amazon AWS}/\emph{Google Cloud Platform} were the top contributors to unsolicited network traffic in \emph{NT-A} and \emph{NT-C}, respectively.

\textbf{Key Takeaway:} This comparison shows that our proactive telescope (\emph{NT-A}) received the most scanning activity from the most distinct sources, many of which were also responsible for the majority of scanning traffic observed in other telescopes. Therefore, in the following sections, we focus on characterizing scanning traffic received by \emph{NT-A}.

\subsection{Unsolicited Traffic Sources by network and traffic type}
\label{sec:results:traffic}

\begin{figure}[h!]
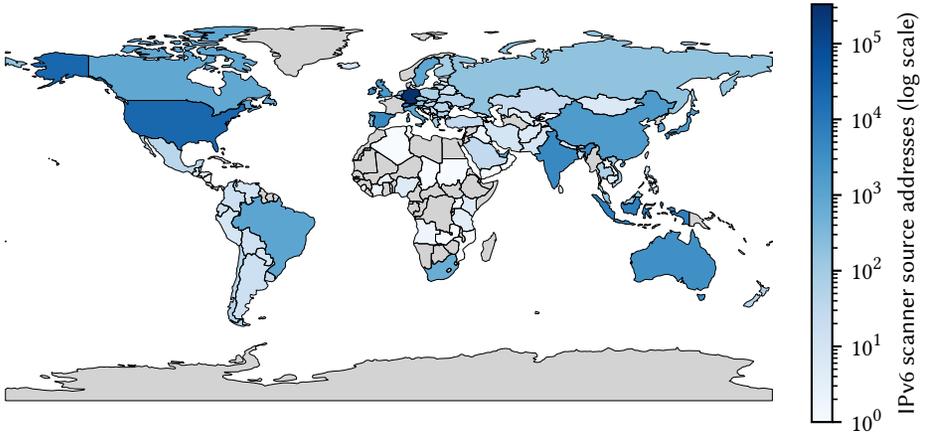

	\centering
\begingroup%
\makeatletter%
%
\makeatother%
\endgroup%
 	\caption{Geographical distribution of /128 scanner source addresses in \echont. We used IPinfo's geolocation database \cite{ipinfogeoip} from April 2024 to infer scanner location from source IPs. Germany is the most prominent source country of scanning due to the vast amount of IP addresses used by \textit{AlphaStrike Labs}.}
	\label{fig:geo_heatmap}
\end{figure}

During our controlled experiments, \echont received 654M unsolicited network
packets from 259k distinct IPv6 addresses spanning over 1.9k unique ASNs.
As the sizes of the subnets that scanners used vary, we aggregated source 
IP addresses using two common prefix lengths (/48, /64). 
\Cref{tab:top5} presents the top five ASNs sourcing such traffic and \Cref{fig:geo_heatmap} shows a geographic distribution of scan sources (a detailed breakdown of top 20 ASNs is shown in \Cref{app:nta_ases}).
\textit{Amazon-02} and \textit{CERNET} accounted11
for 80\% of all observed unsolicited traffic. While 
the traffic volumes from these two source ASNs were comparable,
they differed significantly in the
number of source IP addresses: 44K for
\textit{Amazon-02} and only 46 unique source addresses for \textit{CERNET}.
\begin{table}[h!]
	\small
	\caption{Top 5 ASN sources of unsolicited traffic. (See \Cref{app:nta_ases}, \Cref{tab:top25} for a longer list of top ASNs.)}
	\label{tab:top5}
	\begin{tabular}{cc|c|ccc}
		\toprule
		\multirow{2}{*}{AS name} & \multirow{2}{*}{ASN} & \multirow{2}{*}{Packet count (share \%)} & \multicolumn{3}{c}{Unique sources}\\
		& & & /128 & /64 & /48 \\

		\midrule
		AMAZON-02 & 29014 & 289M (42.6\%) & 44k & 336 & 251 \\
		CNGI-CERNET & 23910 & 265M (39.0\%) & 46 & 4 & 4 \\
		AMAZON-AES & 14618& 33.7M (4.96\%) & 11k & 25 & 15 \\
		TSINGHUA UNI. &45576 & 23.8M (3.51\%) & 5 & 2 & 1 \\
		HURRICANE & 6939 & 12.7M (1.87\%) & 3.5k & 136 & 112 \\

		\bottomrule
	\end{tabular}
\end{table}

\begin{figure}[ht]
	\centering
	\begin{subfigure}[t]{0.495\textwidth}
		\vspace{0pt}
		\centering
		\includegraphics[width=1\textwidth]{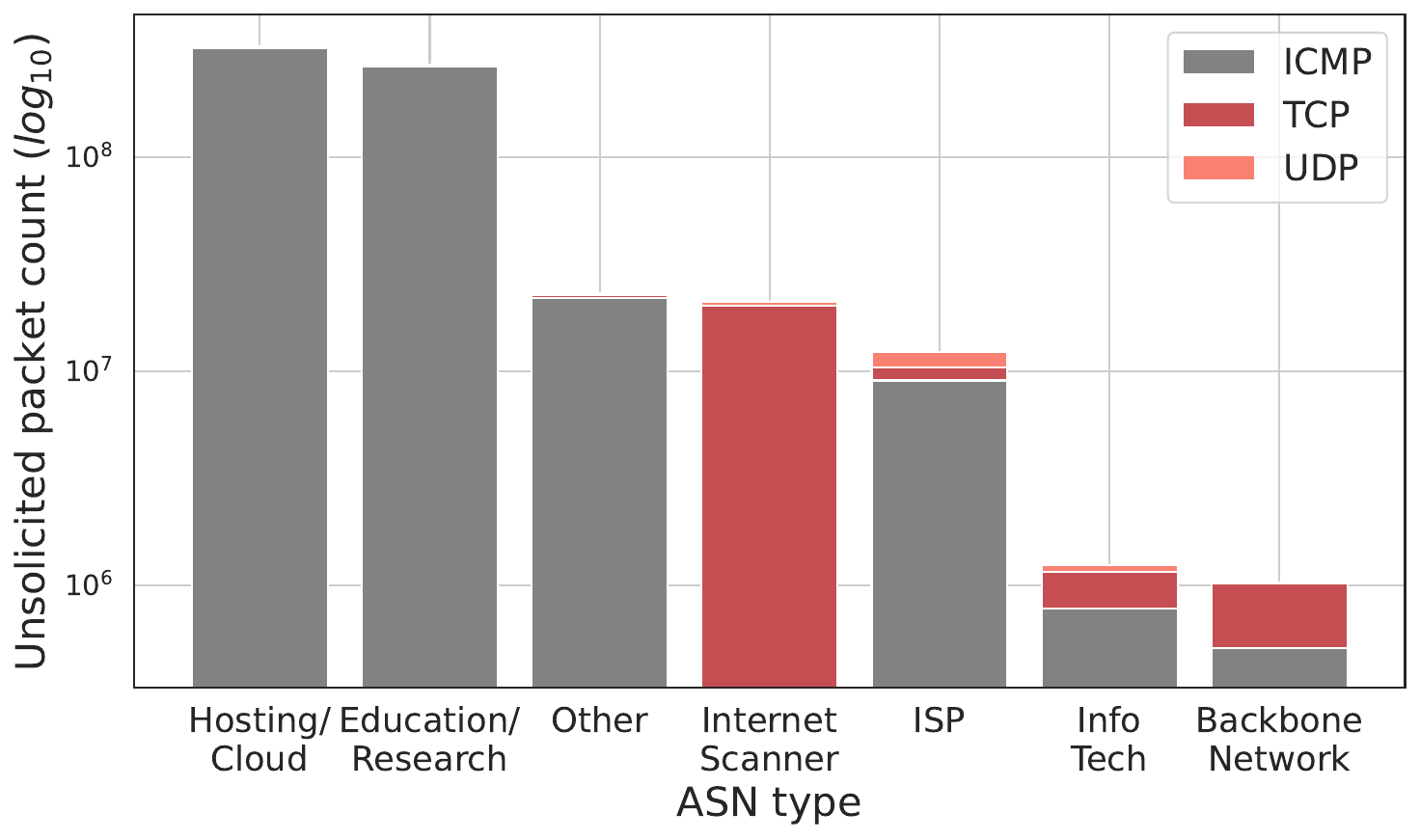}
\caption{Unique packet count categorized by AS types. ICMP dominates 
the traffic, except for \textit{Internet Scanners} which mostly use TCP. }
		\label{fig:scan:traffic}
	\end{subfigure}
	\begin{subfigure}[t]{0.495\textwidth}
		\vspace{0pt}
		\centering
		\includegraphics[width=1\textwidth]{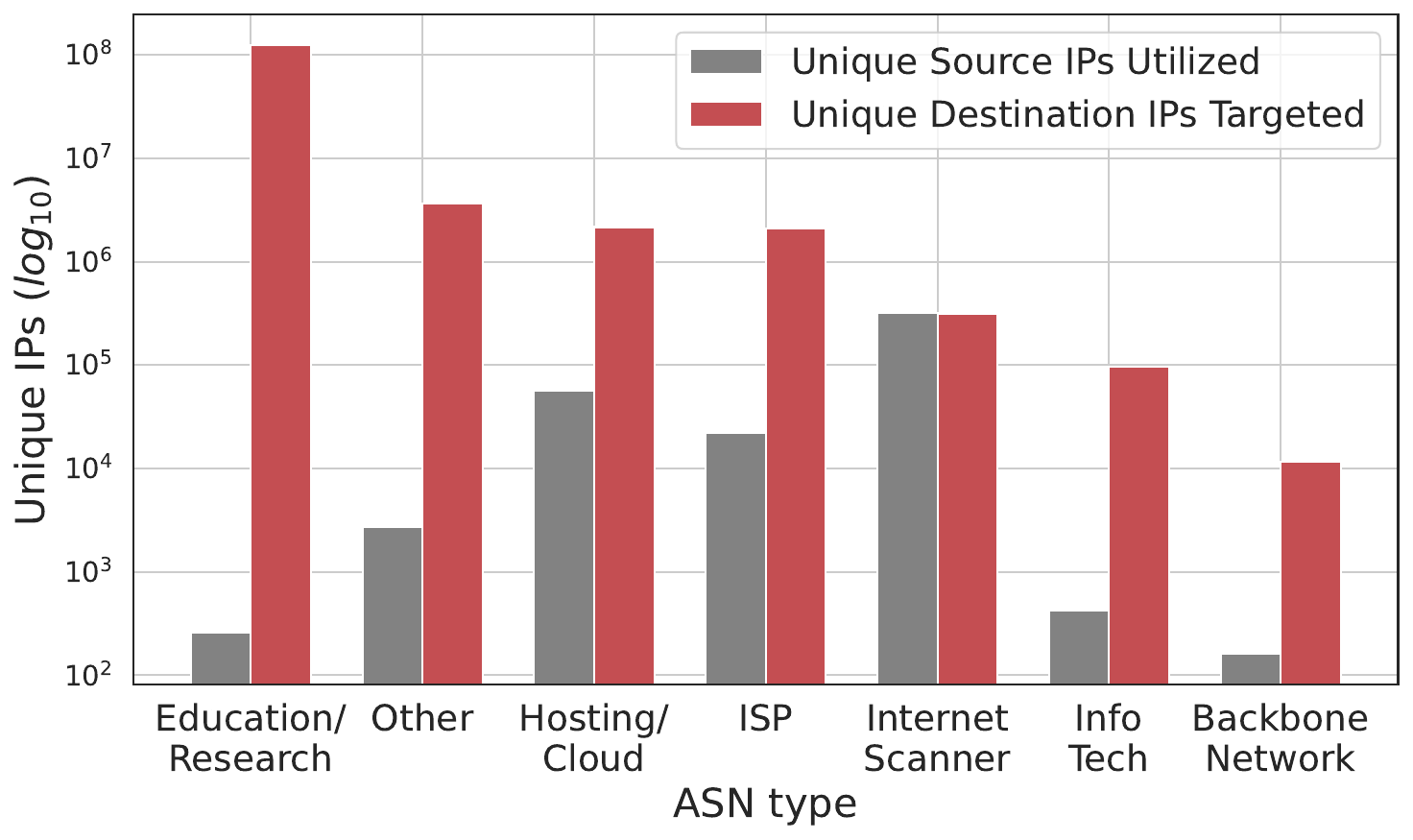}
		\caption{Unique source utilized/destination IPs probed by different AS types. Scanners from R\&E ASes probe an order of magnitude more target IPs than scanners from other ASes.}
		\label{fig:scan:srcdst}
	\end{subfigure}
	\caption{Breakdown of scanner sources by AS type and their proportionate contributions to unique scan sources (/128), unique destinations targeted within our telescope \emph{NT-A} and total number of scanning packets sent broken down by protocol.} \label{figure:asn:nta}
\end{figure}

\Cref{figure:asn:nta} shows the type of all 1.9K source ASes with ASdb \cite{Ziv2021} and statistics on each transport protocol used, proportion of total observed packets, unique 
destinations, and unique sources.
We observed entire IP prefixes and ASes dedicated to IPv6 Internet scanning for various purposes, \eg the Internet Measurement AS \cite{internet19:online}. We manually assigned four such network entities to the \emph{Internet Scanner} category, \eg \textit{AlphaStrike Labs}, \textit{Shadow Server}.

Consistent with previous observations \cite{tanveer2023glowing, richter2022illuminating}, hosting/cloud providers generated the most unsolicited traffic, followed by Research/Education (R\&E) networks. ICMPv6 was the most common protocol, responsible for 91.6\% of all unsolicited traffic.  The \emph{Internet scanner} category contained most (90\%) of the unique source addresses that sent unsolicited traffic to \emph{NT-A}.  These scanners use distributed IP addresses from a covering prefix as large as a /30. The \textit{Internet Scanners} category also presents a different traffic type distribution than other categories, predominantly TCP. R\&E networks probed the most distinct target addresses in \textit{NT-A}, accounting for 95\% of all destinations.

\subsection{Scan Traffic Attraction by Controlled Experiments}
\label{sec:results:effect}
We evaluate the effectiveness of our methods to attract IPv6 scanning traffic. We quantify their impact on traffic volume, scanner diversity, and characterize scanner behavior and target scope. 

\textbf{Impact on scan traffic volume.}
To quantify the increase in scan traffic volume following our controlled experiments, we used ${\Delta_{\mathcal{H}}^{traffic}}$ as defined in 
\S\ref{sec:capture}. \Cref{table:effect_sizes} shows effect sizes for honeyprefixes averaged over all post intervention days. 
Our controlled experiments led to a statistically significant increase in scanning activity across all honeyprefixes
(Tables \ref{table:effect_sizes} and \Ref{fig:small_e}).
We observed the largest ${\Delta_{\mathcal{H}}^{traffic}}$ increase in ${\mathcal{H}^{TLS}_{Tpot1}}$; unsolicited traffic increased by 
224k packets/day for this honeyprefix.
\Cref{fig:small_e} shows the scanning traffic/scanner effect sizes in $\mathcal{H}_{specific}$ subnets. 
The ${\Delta_{\mathcal{H}}^{traffic}}$ sizes in $\mathcal{H}_{specific}$ subnets fall in two distinct classes; ${\Delta_{\mathcal{H}}^{traffic}}$ was below 10k for 75\% of $\mathcal{H}_{specific}$ subnets and above 80k for the rest. The largest ${\Delta_{\mathcal{H}}^{traffic}}$ in $\mathcal{H}_{specific}$ subnet was observed in a covering honeyprefix from which we announced a /61 IPv6 prefix (\S\ref{subsubsec:control}); it received over 10M scanning packets in a single day. However, for the most part, scanning traffic in $\mathcal{H}_{specific}$ subnets was sporadic and we observed no correlation between the prefix length we announced and the scanning traffic we observed to that prefix.

\setlength{\tabcolsep}{2pt}
\begin{table}[h!]
	\centering
	\scriptsize
		\caption{Effect sizes of controlled experiments. Confidence Intervals for traffic rounded to nearest 1000.}
	\label{table:effect_sizes}
	\begin{tabular}{l|*{10}{m{0.9cm}}}
		\toprule
		Honeyprefix & \rotatebox{45}{$\mathcal{H}_{BGP}$} & \rotatebox{45}{$\mathcal{H}_{Alias}$} & \rotatebox{45}{$\mathcal{H}_{TCP}$} & \rotatebox{45}{$\mathcal{H}_{UDP}$} & \rotatebox{45}{$\mathcal{H}_{com}$} & \rotatebox{45}{$\mathcal{H}_{org/net}$} & \rotatebox{45}{$\mathcal{H}_{combined}$} & \rotatebox{45}{$\mathcal{H}_{Tpot1}^{init}$} & \rotatebox{45}{$\mathcal{H}_{Tpot1}^{hitlist}$} & \rotatebox{45}{$\mathcal{H}_{Tpot1}^{TLS}$} \\
		\midrule
		${\Delta_{\mathcal{H}}^{traffic}}$ & 3,865 & 10,670 & 2,525 & 112,000 & 11,903 & 8,159 & 11,493 & 1,115 & 3,457 & 224,176 \\
		{[95\% CI]} & [4k--3k] & [11k--10k] & [3k--3k] & [113k--110k] & [13k--11k] & [8k--8k] & [12k--11k] & [1k--1k] & [8k-- -1k] & [236k--213k] \\
		\midrule
		${\Delta_{\mathcal{H}}^{ASN}}$ & 36 & 25 & 10 & 34 & 35 & 39 & 32 & 26 & 2 & 27 \\
		{[95\% CI]} & [45--28] & [37--13] & [10--10] & [44--25] & [49--21] & [47--32] & [46--18] & [26--25] & [38-- -34] & [117-- -61] \\
		\bottomrule
	\end{tabular}

\end{table}

\textbf{Impact on scanner source diversity.}
To quantify the increase in scanner source diversity, 
we use ${\Delta_{\mathcal{H}}^{ASN}}$ as defined in 
\S\ref{sec:capture} and shown in \Cref{table:effect_sizes}. 
$\mathcal{H}_{org/net}$ had the largest increase in ${\Delta_{\mathcal{H}}^{ASN}}$  (39 source ASNs/day), 
indicating that typical scanners use DNS zone files as seed files to generate scan targets. 
In contrast, most $\mathcal{H}_{specific}$ had an average ${\Delta_{\mathcal{H}}^{ASN}}$ of 1 (not listed in \Cref{table:effect_sizes}), reflecting the limited scanner diversity observed.
This low ${\Delta_{\mathcal{H}}^{ASN}}$, coupled with sporadic scanning traffic, we attribute to the limited propagation of BGP announcements for $\mathcal{H}_{specific}$. Among the 36 public BGP collectors monitored, announcements from $\mathcal{H}_{specific}$ reached only 5 collectors, compared to an average of 28 collectors for other \textit{honeyprefixes}. This result aligns with the established constraint that /48 is the minimum globally routable IPv6 prefix.

\subsubsection{Characterization of Scanner Behavior}
To understand scanner behavior, we reviewed scanning traffic/sources per day following our controlled experiments and uncovered IPv6 scanner behaviors. 

\begin{figure}[h!]
	\centering
	\begin{minipage}{\textwidth}
		\centering
		\includegraphics[width=\textwidth]{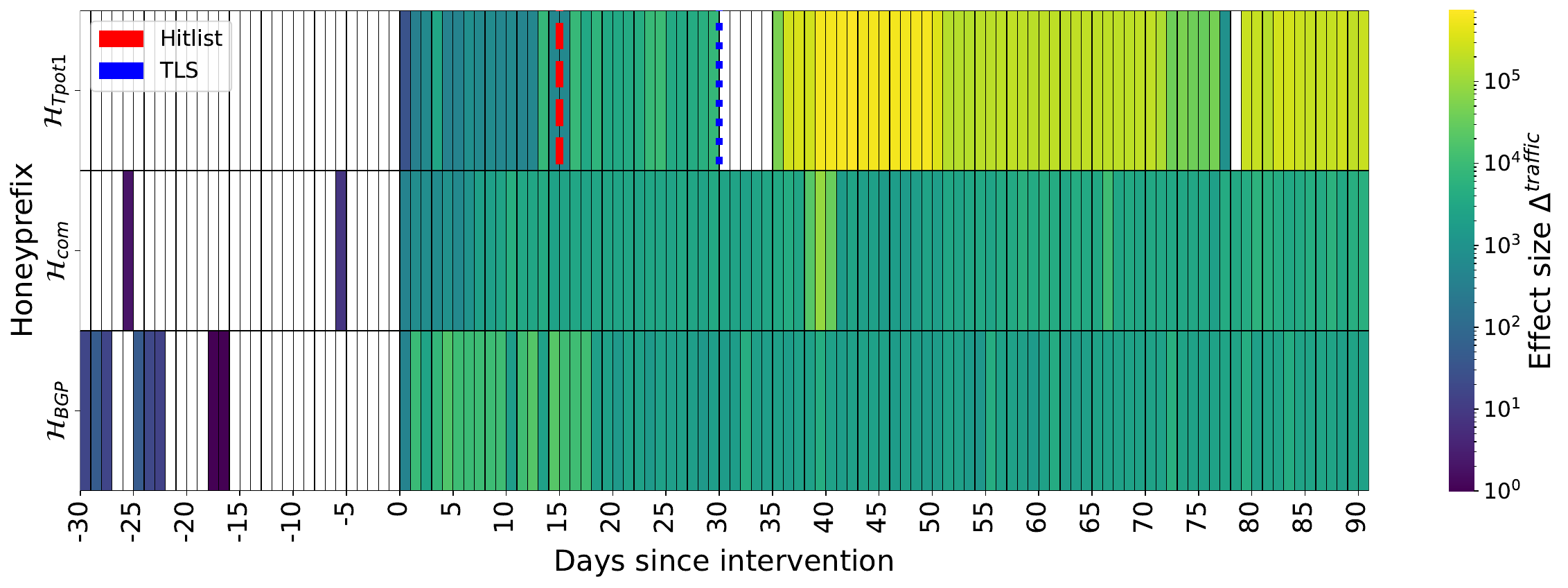}
		\caption{Heatmap of traffic effect size in selected \textit{honeyprefixes}. The figure shows the effect sizes after the initial BGP announcement -- ${\mathcal{I}_{00}}$ -- of 3 \textit{honeyprefixes}. We observe an immediate increase in scanning traffic following the announcement and see traffic increase by an order of magnitude for every additional trigger -- \textit{inclusion in IPv6 hitlist} (red line) and \textit{registering TLS certificates} (blue line).}
		\label{fig:heatmap:traffic}
	\end{minipage}
	
	\vspace{0.5cm} %

	\begin{minipage}{\textwidth}
		\centering
		\includegraphics[width=\textwidth]{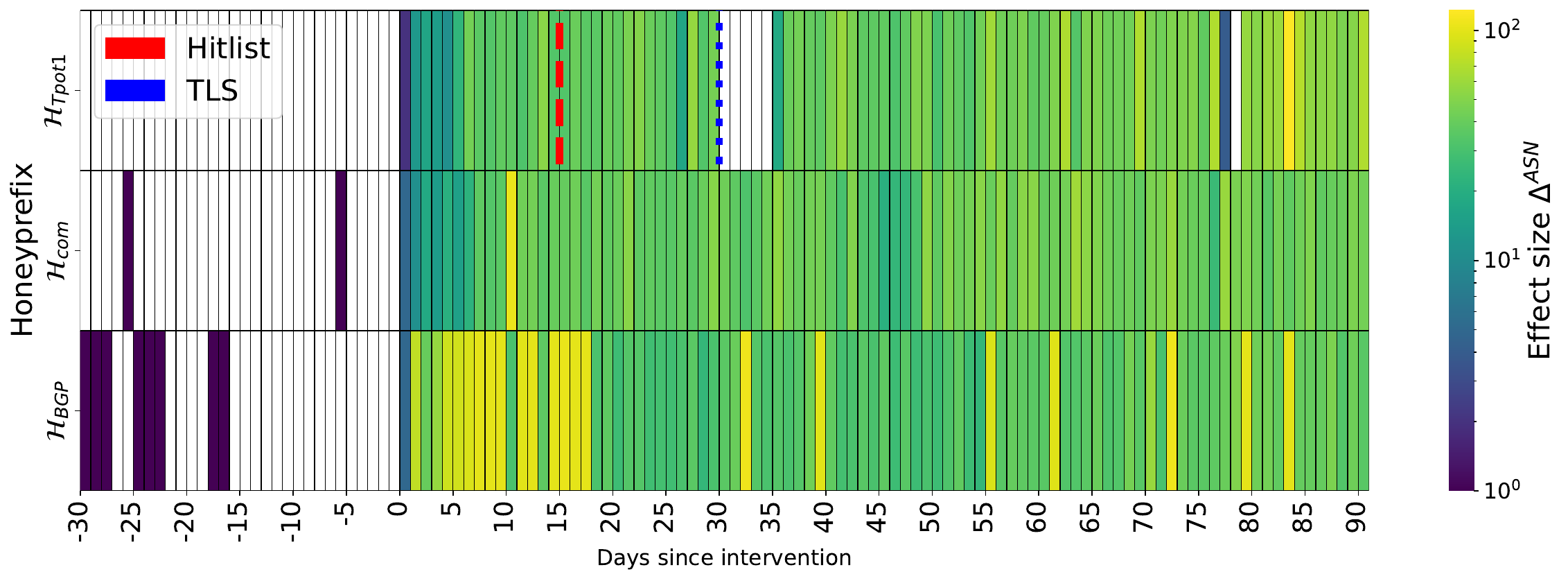}
		\caption{Longitudinal view of per-day traffic and ASN effect sizes in selected \textit{honeyprefixes}. This figure illustrates the effect size in terms of unique source ASNs before and after the initial BGP announcement. Unlike traffic volume, the count of unique source ASes remains consistent, indicating
that scanners consistently probe the \textit{honeyprefixes} but
send less scanning traffic after an initial phase 
(evident in \Cref{fig:heatmap:traffic}).}
		\label{fig:heatmap:asn}
	\end{minipage}
\end{figure}

\textbf{Changes in scanner attention.} We found that the per day difference in scanners/scanning activity fluctuates significantly. Figures \ref{fig:heatmap:traffic} and \ref{fig:heatmap:asn} show that scanner attention increases immediately after our controlled experiment begins, marked by the initial BGP announcement for the honeyprefix. This immediate increase lasts different periods of time for different \textit{honeyprefixes}, suggesting that scanner attention depends on the scanning trigger, \ie which service operates in the \textit{honeyprefix}. \Cref{fig:heatmap:traffic} shows that ${\Delta_{\mathcal{H}}^{traffic}}$ and ${\Delta_{\mathcal{H}}^{ASN}}$ converge to a stable lower value after 15 days for $\mathcal{H}_{specific}$ and 40 days for ${\mathcal{H}^{TLS}_{Tpot1}}$. 
This also explains that despite $\mathcal{H}_{com}$ attracting the most unique scanning source ASes, it received more than an order of magnitude less scanning traffic per day than the interactive honeyprefix $\mathcal{H}_{Tpot1}$. 

\textbf{Scanner response to additional triggers.}
We wanted to understand how scanners react to additional triggers inside a \textit{honeyprefix} a month after the initial BGP announcement.  To this end, we include two additional triggers (inclusion in IPv6 hitlist and registering TLS certificates) in $\mathcal{H}_{Tpot1}$ (red and blue vertical lines in \Cref{fig:heatmap:traffic} and \Cref{fig:heatmap:asn}).  Scanning traffic jumped an order of magnitude with each new trigger, despite the number of source ASes remaining similar. This observation reveals that scanners increase the resources expended on a \textit{honeyprefix} if they detect additional activity in a prefix from disparate data sources.

\textbf{Scanner reaction to trigger retraction.}
To test whether scanners consistently update their sources for seeding IPv6 targets, we retracted the BGP announcements for two of the remaining $\mathcal{H}_{BGP}$ (not shown here) after 4 months. Within hours, the persistent scanning activity diminished to a negligible level, indicating that IPv6 scanners frequently refresh their data sources.

\subsubsection{Characterization of scanner's scope} 
\begin{figure}[h!]
	\centering
	\begin{minipage}{.485\textwidth}
		\centering
		\includegraphics[width=\linewidth]{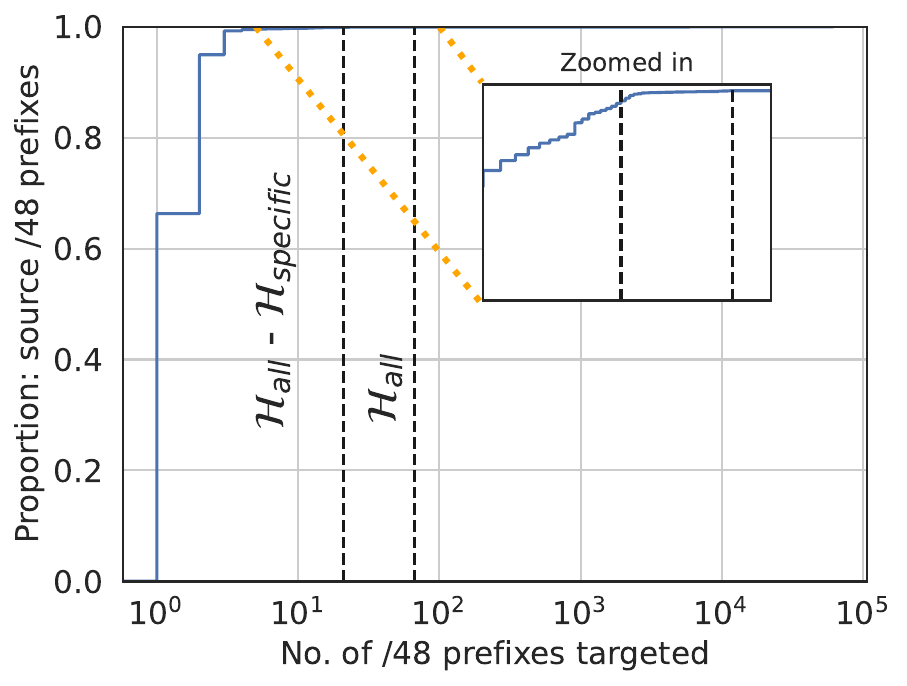}
		\caption{Number of /48 prefixes targeted by observed scanners. 98.4\% of scan traffic we received in our proactive telescope \emph{NT-A}, was directed to one of our \textit{honeyprefixes}, with 99.9\% of this traffic targeting \textit{honeyprefixes} other than $\mathcal{H}_{specific}$.  That is
most IPv6 scanners stay within the prefix address scope 
of the \textit{honeyprefix}}.
		\label{fig:scanner_scope}
	\end{minipage}
	\hfill
	\begin{minipage}{.485\textwidth}
			\includegraphics[width=\linewidth]{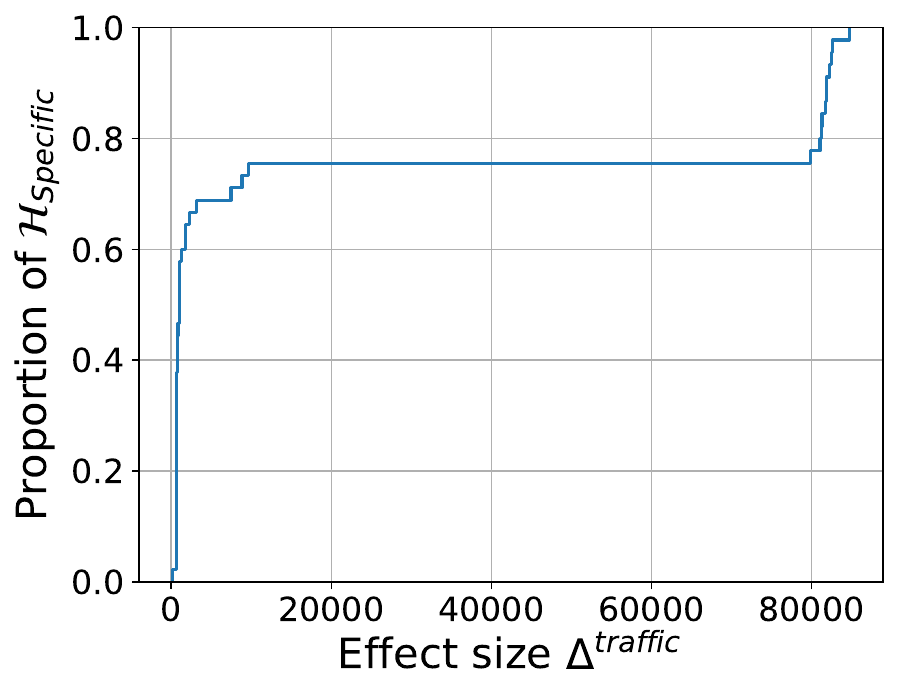}
		\caption{$\Delta^{traffic}$ effect size for $\mathcal{H}_{specific}$ subnets. We observe that most (75\%) $\mathcal{H}_{specific}$ \textit{honeyprefixes} receive <10k scanning packets and the remaining receive > 80k. This is a result of prefixes longer than /48 not propagating through BGP. \label{fig:small_e}}
	\end{minipage}
\end{figure}

We defined IP prefix boundaries for our controlled experiments by announcing a /48 for each of our \textit{honeyprefixes} from within our telescope's /32 covering prefix. To analyze whether the IPv6 scanners stayed within the address scope of the honeyprefixes, we analyzed the number of /48 prefixes that scanners probed (\Cref{fig:scanner_scope}). 95\% of scanners probed only two /48 prefixes, 99.92\% probed $<$11 prefixes (excluding $\mathcal{H}_{specific}$ subnets), and 99.97\% probed $<$27 (the total number of honeyprefixes we deployed in our experiment). Only 55 of 191k scanner source IPs scanned beyond this scope, with one probing 61.5k of 65k /48 prefixes. Non-honeyprefix traffic constituted a meager 1.6\% of total unsolicited traffic we received in the covering /32 prefix, half of which targeted the first 16 /48 prefixes (xxxx:xxxx:0000::/48 to xxxx:xxxx:000f::/48), with the rest distributed across the remaining subnets. This finding highlights that \textbf{1)} /48 is an adequate prefix size to host such controlled experiments and \textbf{2)} most IPv6 scanners scan within the prefix address scope of the honeyprefix.

\textbf{Key Takeaway:} Our findings offer foundational guidance for future IPv6 telescope design. A /48 prefix is essential -- not only because it is the minimum globally routable prefix, but also because it allows precise attribution of scanning activity; broader covering prefixes can cause collateral scanning spillover. We observe that scanners strictly confine their probing to the announced /48, reinforcing its utility for controlled experiments. Operationally, resource-constrained deployments should prioritize exposing addresses via TLS certificates and deploying both high and low interaction honeypots. These mechanisms significantly amplify scanner engagement, and hosting multiple services within the same /48 can compound visibility—provided the prefix remains consistently announced via BGP.

\subsection{Characterizing scanning strategies} 
\label{sec:results:scanstrategy}
\begin{figure}[h!]
	\centering
	\begin{subfigure}[b]{.9\textwidth}
		\centering
		\includegraphics[trim={0 3.5cm 0 0}, clip, width=0.9\textwidth, height=0.25\textheight]{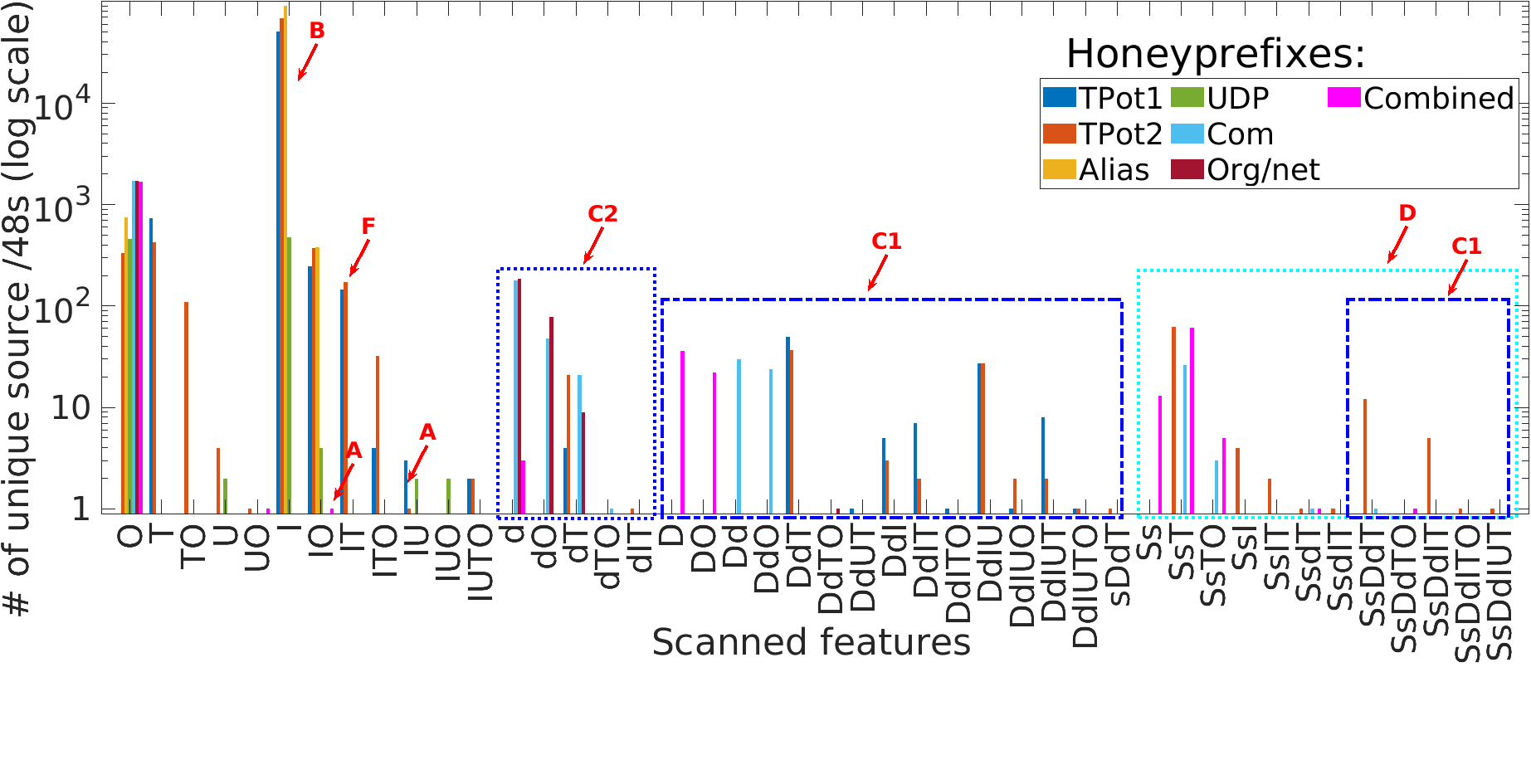}
		\caption{Many sources discovered our honeyprefixes using domain names and CTLog, not the IPv6 hitlist.} \label{fig:scanfeat:nonhitlist}
	\end{subfigure}
	\begin{subfigure}[b]{.9\textwidth}
		\centering
		\includegraphics[trim={0 2.5cm 0 0}, clip, width=0.9\textwidth, height=0.25\textheight]{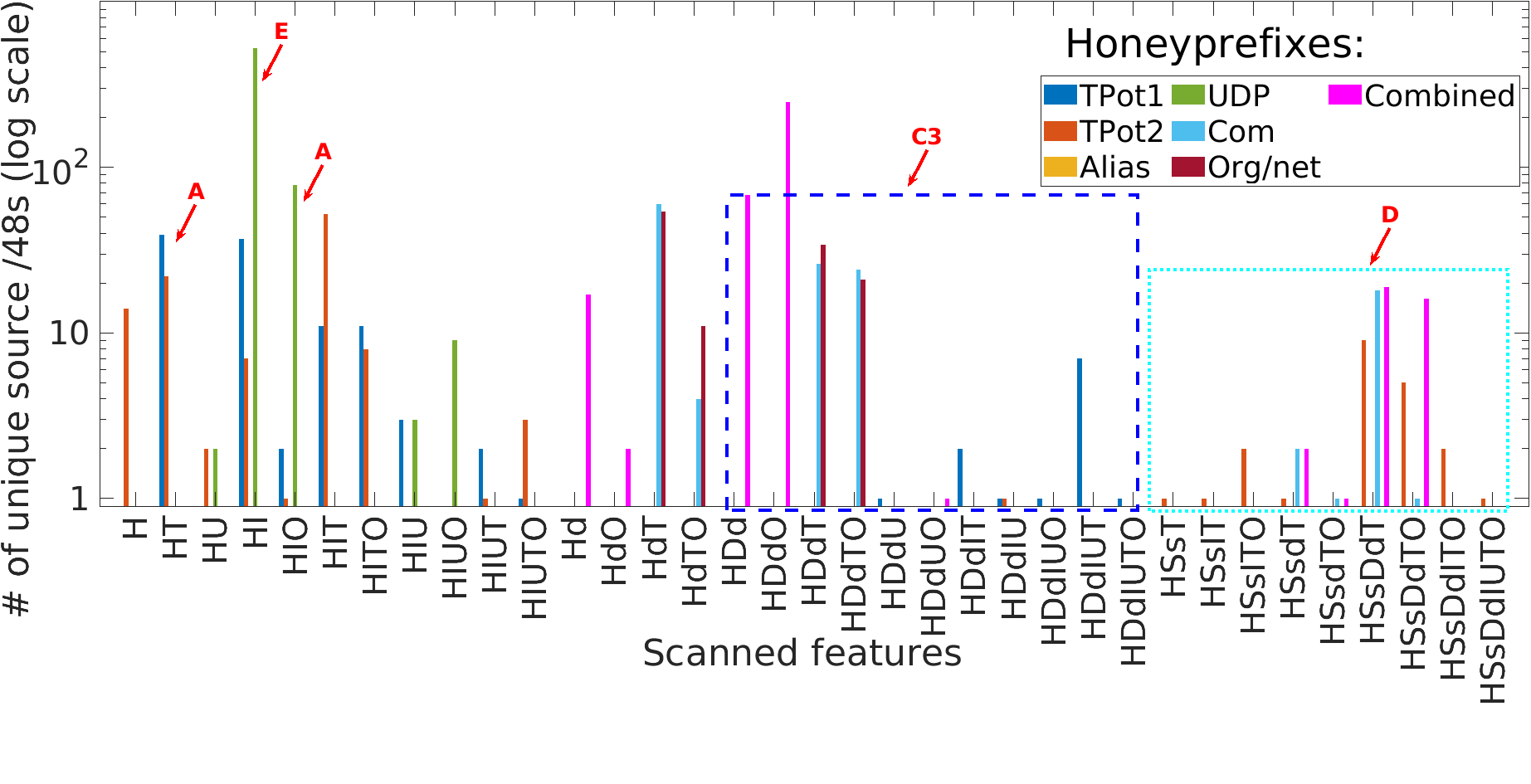}
		\caption{Manually inserted addresses in the TPots reveals that scanners use the IPv6 hitlist to populate target lists.} \label{fig:scanfeat:hitlist}
	\end{subfigure}
	\caption{Combination of tactics adopted by scanning traffic sources, grouped by /48. Each colored bar shows the number of scanners using the same probing strategy against honeyprefixes. \emph{X}-axis labels represent combinations of features we introduced (\S\ref{sec:expr:protelescope:feat}). \feat{O} represents scanners probing non-responsive protocols or ports. Red arrows highlight the findings described in \S\ref{sec:results:scanstrategy}. We exclude results from $\mathcal{H}_{TCP}$ because its /48 prefix was not successfully announced via BGP, leading to only limited probing activity --- specifically, traffic from just seven /48s across five ASes.} \label{fig:scanfeat}
\end{figure}

The unique characteristics of each honeyprefix suggest which services and protocols were targeted by IPv6 scanners. \Cref{fig:scanfeat} shows the features that scanning /48 subnets probed. Each \emph{x}-axis label represents a combination of features in nine responsive honeyprefixes that scanners hit. We matched probe packets with honeyprefixes' features by protocols, (\eg ICMP and TCP/UDP destination ports), the destination IPs and the probing time.   We used the probing time to distinguish features that shared the same protocols and destination addresses. For example, both  (sub-)domain names (\feat{S} and \feat{D} in \Cref{fig:scanfeat}) and their corresponding TLS certificates (\feat{s} and \feat{d} in \Cref{fig:scanfeat}) resolve to web services at the same IPs. 
We labeled scanning attempts targeted to domain names or TLS certificates 
for probes received before or after certificate issuance, respectively.

\textbf{ICMP probing.} As the entire $\mathcal{H}_{TPot1}$, $\mathcal{H}_{TPot2}$, and $\mathcal{H}_{Alias}$ were responsive to ICMP, over 50k unique sources probed addresses in these three honeyprefixes \emph{solely} using ICMP pings (\finding{B} in \Cref{fig:scanfeat:nonhitlist} and \ref{fig:scanfeat:hitlist}). About 400 sources probed $\mathcal{H}_{UDP}$ and $\mathcal{H}_{RDNS}$ solely with ICMP, despite both honeyprefixes having only 3 addresses responding to ICMP. All these sources targeted only the first address (\texttt{::1}) of the subnet, although two other addresses were active. Two sources discovered the random address for which we enabled ICMP response in $\mathcal{H}_{Combined}$. They also scanned UDP ports and other non-responsive parts of the subnet (magenta bars in ``IU'' and ``IO'', \finding{A} in \Cref{fig:scanfeat:nonhitlist}).

\textbf{Domain registration.} Many sources used the root AAAA DNS records to compile target lists (\ie \feat{D}). All honeyprefixes with domain names ($\mathcal{H}_{Com}$, $\mathcal{H}_{Org/net}$,  $\mathcal{H}_{Combined}$, $\mathcal{H}_{TPot1}$, and $\mathcal{H}_{TPot2}$) showed evidences of scanning targeted to the DNS-mapped IPs (\feat{D} in \emph{x}-axis, indicated by \finding{C1} in \Cref{fig:scanfeat:nonhitlist} and \ref{fig:scanfeat:hitlist}). 
These scanners likely learned the names (and corresponding addresses) from published zone files, as there was a month-long gap between our domain registration (Sept. 19, 2023) and the associated IP appearing on the IPv6 hitlist (Oct. 23, 2023).

\textbf{Subdomain names and TLS certificates.} No sources could detect common subdomains without TLS certificates (\ie \feat{s} always came with \feat{S} in \Cref{fig:scanfeat}, indicated by \finding{D}). 
Even for root domain names, $\sim$300 scanners observed by $\mathcal{H}_{Com}$, $\mathcal{H}_{Org/net}$ only probed the honeyprefixes after the issuance of TLS certificates (\finding{C2} in \Cref{fig:scanfeat:nonhitlist}).
Scanners quickly reacted to new TLS certificates. The first scanner from DigitalOcean arrived 7 seconds after certificate issuance, evidence that Certificate Transparency logs \cite{ctlog} were the source of this target list.

\textbf{IPv6 Hitlist.} Manual addition of hitlist entries for $\mathcal{H}_{TPot1}$ and $\mathcal{H}_{Tpot2}$ let us isolate the effect of using hitlists. 115 and 111 sources probed addresses in $\mathcal{H}_{TPot1}$ and $\mathcal{H}_{Tpot2}$ (and corresponding protocols) specified in the hitlists (blue/orange bars in \Cref{fig:scanfeat:hitlist}), respectively. 
We label scanners with \feat{H} that probed addresses that the hitlist automatically discovered. Although we could not confirm that scanners used the IPv6 hitlist, these IPs saw traffic levels similar to the honeypots. Over 600 sources pinged $\mathcal{H}_{UDP}$, which had one responsive IP address that we manually inserted into the IPv6 responsive addresses hitlist (\finding{E} in \Cref{fig:scanfeat:hitlist}).

\textbf{Scanning tactics.} Most scanners used only ICMP to measure the liveness of the networks. Combining TCP and ICMP probing was a common strategy (\finding{F} in \Cref{fig:scanfeat:nonhitlist}). Scanners probed our honeyprefixes with multiple protocols, particularly for the high-interaction honeypots. Ten scanners probed our honeypots with all ICMP, TCP, and UDP protocols.

\textbf{Key Takeaway:} %
All features, except subdomains with TLS certificates, effectively attracted scanners. Most scanners use ICMP to test connectivity, but our proactive telescope can detect sophisticated, potentially malicious scanners that integrated multiple data sources to explore live hosts and services in the targeted networks.

\section{Discussion}\label{sec:limitations}
We discuss limitations and future direction of this work.

\textbf{Generalizability of Our Findings.} 
While our study collected one of the largest known datasets of unsolicited IPv6 traffic, our findings may not generalize to all networks or geographic locations. This limitation has been observed in IPv4 network telescopes, where researchers found that scanning activity can be localized \cite{Richter2019akamaiscan}, introducing biases across different vantage points \cite{wan2020origin}. Additionally, certain network types—such as public cloud infrastructures—have been found to attract more diverse and aggressive scanning behavior than others \cite{pauley2023dscope}. We expect similar patterns to emerge in IPv6 scanning.
Moreover, IPv6 scanning is heavily influenced by the evolution of target generation algorithms, whose effectiveness depends on the seeds and datasets they use \cite{Williams2024seeds}. Some scanners may rely on signals not included in our experiments, potentially resulting in false negatives.

\textbf{Misattribution of Scanning.} We took careful steps to accurately attribute unsolicited traffic to each of our experiments—for example, by precisely timing and placing /48 \textit{honeyprefixes} within the /32 covering prefix. However, our methodology relies on external datasets and mechanisms that may introduce inaccuracies. For instance, our manual inspection revealed misclassifications in ASdb \cite{Ziv2021}. Another source of errors could be related to routing security. Due to the increasing adoption of RPKI, the upstream provider for \ucsdnt would reject BGP announcements of honeyprefixes until the corresponding Route Origin Authorization (ROA) records were successfully registered. As a result, we cannot rule out the possibility that scanners used ROA data, rather than BGP propagation, to infer the presence and location of honeyprefixes.

\textbf{Implications for Operational Security.} 
Traditional IP blocklists and threat intelligence feeds (\eg \cite{abuseip,firehol}) rely on stable, individual IPs to identify malicious actors. However, in IPv6, network operators assign  large prefixes (\eg /48 to /112) to end-hosts \cite{rfc6177}, which complicates attribution and mitigation. 
Overblocking causes collateral damage, while underblocking enables evasion via address rotation.
Accurate mapping of prefixes to hosts requires (1) insights into real-world IPv6 allocation and rotation practices \cite{padmanabhan2020dynamips,rye2021follow}, and (2) large-scale measurement of scanner behavior. Our study contributes to the latter. For example, we identified scanners operating across /32 prefixes. 
IETF drafts have offered recommendations for IPv6 blocklisting \cite{draft-ietf-opsec-ipv6-addressing} and to specify end-site prefix lengths of their networks \cite{draft-ietf-opsawg-prefix-lengths}.
Our findings can empirically ground such operational guidance as we find evidence of IPv6 scanners utilizing as large as a /29 prefix and as small as a /112 prefix to conduct scanning. Without end-site prefix length specification, developing high accuracy IP/prefix based blocklisting tools will be a complex challenge. 
We plan to integrate datasets collected and tools developed in this work for (1) creating effective IPv6 defenses \eg commercially deployed IPv6 only telescopes and (2) helping existing solutions/suggestions to improve \cite{Implicat64:online,TheNoCom78:online,SixIntWo76:online}.

The deployment of IPv4 network telescopes is often constrained by address exhaustion, a limitation that IPv6 does not face.
However, existing IPv6-focused platforms that could support such efforts (\eg Honeydv6 and 6Guard \cite{GitHubCh40:online,GitHubmz37:online}) are outdated or no longer maintained. 
Our open-source tools enable network operators to deploy IPv6 proactive telescopes within their own networks to gather network-specific threat intelligence. 
We plan to improve the scalability and reliability of our software and deployment framework, particularly for high-interaction honeypots, to further strengthen the security of IPv6 networks.

Small networks may lack the resources or autonomy to implement all the features we explored. Operators might have limited control over BGP announcements or 
hold only a single /48 prefix. In such cases, they must balance capability with resource constraints. For instance, rather than deploying multiple honeyprefixes, operators can focus their efforts by placing honeypots, registering domain names, and issuing TLS certificates near the beginning of their assigned address block, where scanners are most likely to initiate probing, to maximize
visibility of scanning activities.

\section{Conclusion}
\label{sec:conclude}
In this work, we designed and developed a proactive telescope --- a reproducible system to capture a broader and more representative range of IPv6 scanner behaviors within an ISP network. Our deployment collected the largest IPv6 telescope dataset to date, in terms of both address space and traffic volume, enabling us to investigate the ecosystem of IPv6 unsolicited traffic.
We carefully analyzed the (in)effectiveness of common network features and public datasets that scanners might leverage for IPv6 network discovery. Our longitudinal measurement data further revealed the strategies scanners use to integrate target addresses and explore network services.
These insights are valuable for network operators and researchers in developing effective tools to mitigate potential threats in IPv6 networks. 
We will publish our code and deployment instructions to support reproducibility this work.
\begin{acks}
We thank our shepherd and the anonymous reviewers for their insightful comments. 
This material is based on research sponsored by the National Science Foundation (NSF) grants NSF OAC-2131987, OAC-2319959, and OAC-2450552. The views and conclusions contained herein are those of the authors and should not be interpreted as necessarily representing the official policies or endorsements, either expressed or implied, of the funding agencies.
\end{acks}

\clearpage \bibliographystyle{plain}
\bibliography{telescope}
\appendix
\newpage
\section{Ethics}

This work does not raise ethical concerns.
\section{T-Pot Infrastructure}
\label{app:tpot}

\begin{minipage}{\textwidth}
\begin{minipage}[c]{0.47\textwidth}
	\centering
	\includegraphics[width=\textwidth]{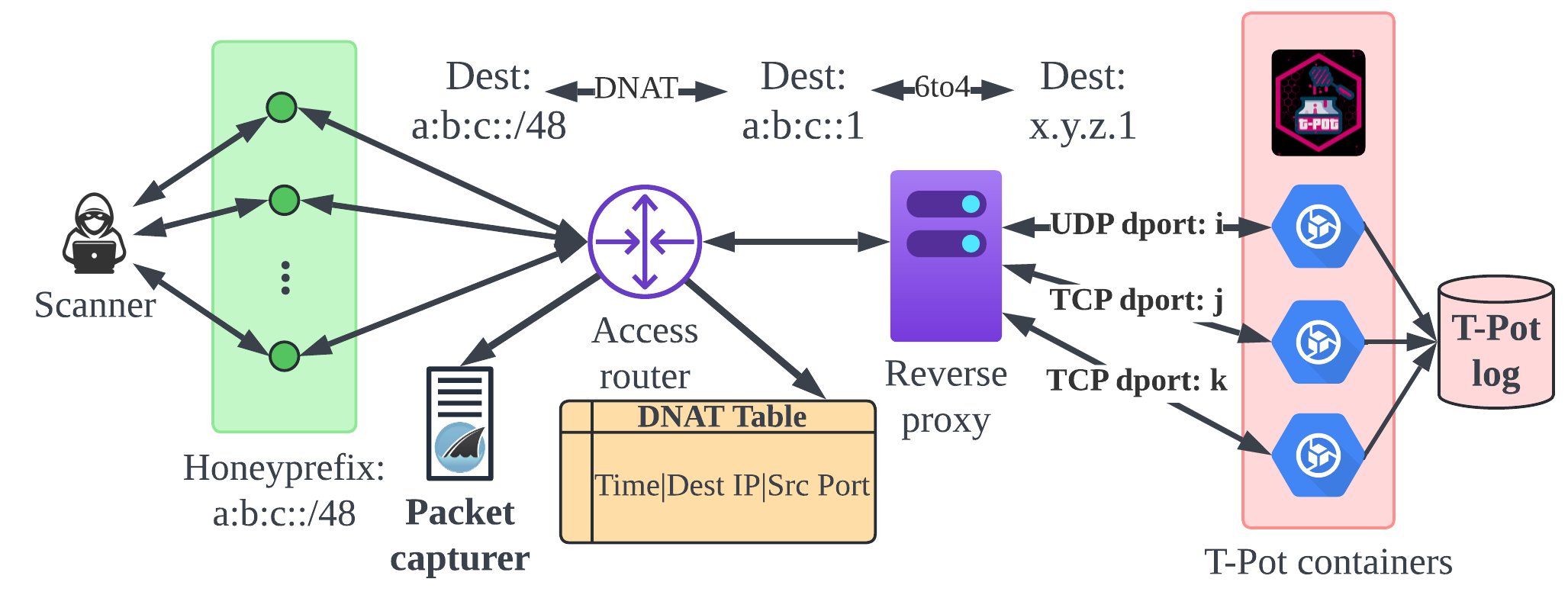}
	\captionof{figure}{Overview of IPv6-enabled T-Pot infrastructure.} \label{fig:tpotinfra}
\end{minipage}
\hfill
\begin{minipage}[b]{0.51\textwidth}
	\tiny
	\centering
	\captionof{table}{Honeypot containers we deployed in our T-Pot instance and the corresponding ports.}
	\label{table:docker_ports}
	\begin{tabular}{lp{3cm}cc}
		\toprule
		\textbf{Honeypots} & \textbf{Protocol (destination ports)} & $\mathcal{H}_{TPot1}$ & $\mathcal{H}_{TPot2}$ \\ 
		\midrule
		cowrie & TCP (22-23) & \cmark & \xmark\\ 
		mailoney & TCP (25) & \cmark  & \cmark\\ 
		snare & TCP (80) & \cmark & \cmark\\
		citrixhoneypot & TCP (443) & \cmark & \cmark\\
		ciscoasa & UDP (5000), TCP(8443) & \cmark & \cmark \\ 
		redishoneypot4 & TCP (6379) & \cmark & \xmark\\ 
		adbhoney & TCP (5555) & \cmark & \cmark\\
		sentrypeer & UDP (5060) &\xmark & \cmark\\
		dionaea & TCP (20-21, 42, 81, 135, 443, 445, 1433, 1723, 1883, 3306, 27017), UDP (69)& \cmark & \cmark\\ 
		ddospot & UDP (19, 53, 123, 161, 1900) & \cmark & \cmark\\ 
		conpot\_kamstrup\_382 & TCP (1025, 50100) & \xmark & \cmark\\ 
		elasticpot & TCP (9200) & \xmark & \cmark\\ 
		dicompot & TCP (11112) & \xmark & \cmark\\ 
		\bottomrule
	\end{tabular}
\end{minipage}
\end{minipage}

\section{Breakdown of Scanner Sources in CDN by Country and Network Type}
\label{app:cdn}

Also of interest is the country of origin and network type from which the scans are initiated. 
Table~\ref{tab:sourceases} shows the top 20 ASes, ordered by number of packets.  For countries, United States and China dominate.
Two of the ASes belong to cybersecurity companies.  
If the scanner has ill-intent, they could likely use cloud service providers or datacenters (though of course others would also be using these
platforms) and these platforms are the most popular.
Compared to an earlier study by Richter \textit{et. al}~\cite{richter2022illuminating}, covering 15 months starting January 2021, the scan traffic reported here is much more dispersed.  
In Table~\ref{tab:sourceases} the top AS accounted for 18\% of the packets across three /64's, whereas earlier, the top three /64's accounted for 87\% of the packets.

\vspace{1em}
\begin{minipage}{\textwidth}
	\begin{minipage}[c]{0.49\textwidth}
		\centering
		\includegraphics[width=0.98\linewidth]{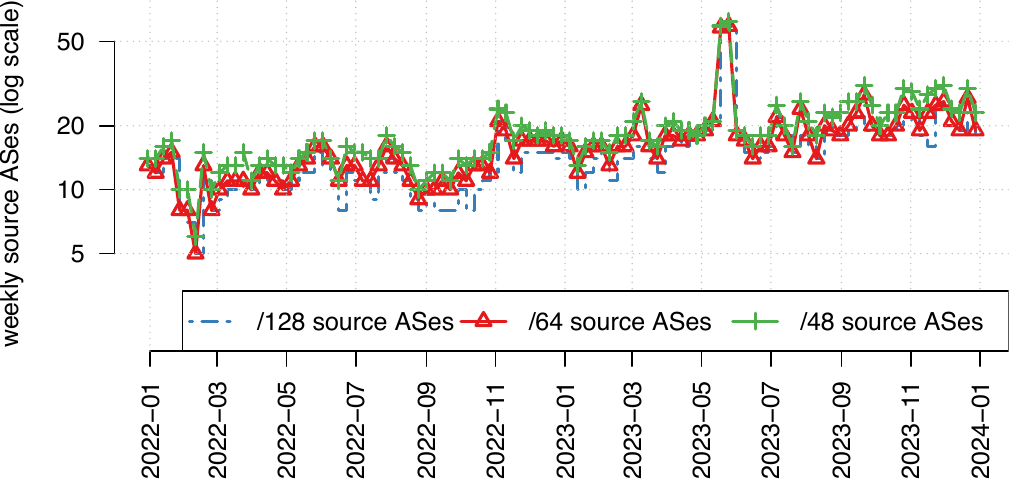}
		\captionof{figure}{Weekly number of source ASes of the IPv6 scans at the CDN.}
		\label{fig:source_ases_over_time}
	\end{minipage}
	\begin{minipage}[b]{0.48\textwidth}
		\scriptsize
		\centering
		\captionof{table}{Top 20 source ASes at the CDN by scan packets over the entire measurement window (packets shown for /64 source aggregation).}
		\label{tab:sourceases}
		\begin{tabular}{rlrrrr}
			\toprule
			\multicolumn{3}{c}{} & \multicolumn{3}{c}{\textbf{Scan sources}} \\
			\cmidrule{4-6}
			\textbf{Rank} & \textbf{AS type} & \textbf{Packets} & \textbf{/48s} & \textbf{/64s} & \textbf{/128s} \\
			\midrule
			\#1 & Transit (global) & 4.68B (17.6\%) & 1 & 3 & 2745\\
			\#2 & Datacenter (CN) & 4.08B (15.4\%) & 10 & 12 & 45\\
			\#3 & Cybersecurity (US) & 3.74B (14.1\%) & 7 & 7 & 367\\
			\#4 & Datacenter (US) & 3.17B (12.0\%) & 1 & 1 & 11\\
			\#5 & Cloud (CN) & 2.60B (9.8\%) & 15 & 17 & 310\\
			\#6 & Cloud (CN) & 2.42B (9.1\%) & 6 & 7 & 36\\
			\#7 & Datacenter (CN) & 1.72B (6.5\%) & 2 & 2 & 11\\
			\#8 & Cloud (US/global) & 899M (3.4\%) & 35 & 43 & 3312\\
			\#9 & Cloud (US/global) & 833M (3.1\%) & 4 & 4 & 53\\
			\#10 & Datacenter (CN) & 609M (2.3\%) & 1 & 1 & 4\\
			\#11 & Cloud (US/global) & 533M (2.0\%) & 12 & 12 & 2277\\
			\#12 & Cloud (US/global) & 392M (1.5\%) & 12 & 19 & 4475\\
			\#13 & Cloud (US/global) & 360M (1.4\%) & 22 & 22 & 41\\
			\#14 & Cloud (US/global) & 228M (0.9\%) & 7 & 7 & 21\\
			\#15 & Cybersecurity (US) & 91M (0.3\%) & 2 & 2 & 198\\
			\#16 & Datacenter (CN) & 44M (0.2\%) & 32 & 138 & 142\\
			\#17 & Cloud (US) & 28M ($\leq$0.1\%) & 1 & 1 & 2\\
			\#18 & University (CN) & 20M ($\leq$0.1\%) & 1 & 2 & 2\\
			\#19 & Datacenter (CA) & 14M ($\leq$0.1\%) & 1 & 1 & 1\\
			\#20 & Research (DE) & 14M ($\leq$0.1\%) & 1 & 1 & 1\\
			\bottomrule
		\end{tabular}
	\end{minipage}
\end{minipage}

\Cref{fig:source_ases_over_time} shows the weekly number of ASes sending IPv6 scan packets. Like the number of source addresses and packets, the number of scanning ASes grows steadily over time (cf. \Cref{sec:newintro}).

\section{Technical Details of Twinklenet}%
\label{app:twinklenet}
We implemented Twinklenet in Go, using Berkeley Packet Filters (BPF) and PCAPGO \cite{pcapgo}
to filter, capture, and respond to incoming packets. 
Twinklenet leverages raw socket to overrides the system's default TCP/IP stack and routing table, enabling it to return the responses to the sender, emulating live hosts within the honeyprefixes. \Cref{tab:twinklenetfunction} lists the services Twinklenet emulates.
\begin{table*}[h!]%
	\centering
	\small
	\caption{Protocols and interactions supported by Twinklenet.}
	\label{tab:twinklenetfunction}
	\resizebox{\textwidth}{!}{%
		\begin{tabular}{cll}
			\toprule
			Protocols & Request & Response(s) \\
			\midrule
			ICMP/ICMPv6 & ICMP/ICMPv6 Echo request & ICMP/ICMPv6 Echo reply \\ \hline
			\multirow{2}{*}{TCP} & TCP SYN to an open port & Complete three-way handshake and close the connection with FIN\\
			& Other TCP packet to an open port& TCP RST\\ \hline
			NTP (over UDP) & Any client NTP packet & NTP Kiss-of-Death packet (Reference Identifier=\texttt{DENY})\\ \hline 
			DNS (over UDP) & Any DNS query & DNS response with response code \texttt{SERVFAIL}\\ 
			\bottomrule
		\end{tabular}
	}%
\end{table*}%

\section{Honeyprefix Location in \echont}
\label{app:honey_location}
\begin{figure}[h!]
	\centering
	\includegraphics[width=.5\textwidth]{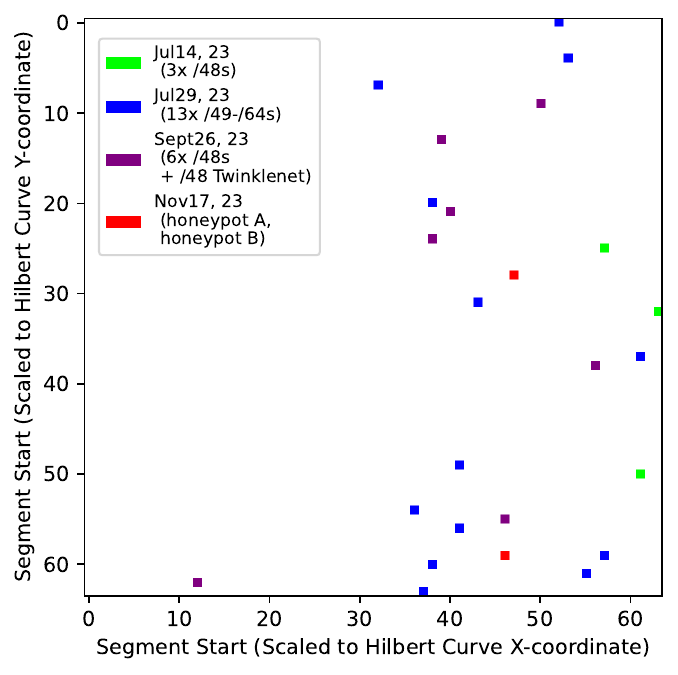}
	\caption{Hilbert map of the /32 IPv6 address space of our telescope in the ISP's network. We distributed our /48 honeyprefixes randomly across the upper half of unused portion of the /32. The colors of the pixels show different phrases of honeyprefix deployment.}
	\label{fig:prefix_announce_map}
\end{figure}

\clearpage

\section{Sources of Scan Traffic Observed in \echont}
\label{app:nta_ases}
\begin{table}[h!]
	\small
	\caption{Top 20 ASN sources of unsolicited traffic in \echont (continued from \Cref{tab:top5}).} \label{tab:top25}
	\begin{tabular}{rllcrrr}
		\toprule
		& & & & \multicolumn{3}{c}{\bf Unique sources}\\
		\cmidrule{5-7}
		\bf Rank& \bf AS name& \bf ASN& \bf Packet count (share \%) & \bf /128 & \bf /64 & \bf /48 \\
		\midrule
		\multicolumn{7}{c}{\ldots \textit{(see \Cref{tab:top5}})}\\
\#6 &			BJ-GUANGHUAN-AP &	55960 &		7.1M (1.04\%)&695&	8&	4	\\
\#7 &			ALIBABA-CN-NET &	45102 &		7.0M (1.03\%)&17&	9&	8	\\
\#8 &			INTERNET-MEASUREMENT &	211298 &	6.9M (1.01\%)&511&	14&	14	\\
\#9 &			AKAMAI-AMS &		33905 &		5.5M (0.81\%)&3&	3&	3	\\
\#10 &			PONYNET &		53667 &		4.3M (0.63\%)&10&	10&	4	\\
\#11 &			UONET &			3582 &		3.4M (0.50\%)&2&	2&	2	\\
\#12 &			WESTCLOUDDATA &		135629 &	3.4M (0.50\%)&374&	3&	1	\\
\#13 &			NEXTLAYER &		1764 &		1.9M (0.28\%)&11&	11&	7	\\
\#14 &			CHOOPA &		20473 &		1.0M (0.15\%)&207&	207&	54	\\
\#15 &			LEITWERT-RESEARCH &	29108 &		1.0M (0.15\%)&11&	11&	11	\\
\#16 &			BCIX &			62193 &		1.0M (0.15\%)&1&	1&	1	\\
\#17 &			MWN &			12816 &		880K (0.13\%)&30&	2&	2	\\
\#18 &			AKAMAI-ASN1 &		20940 &		779K (0.11\%)&6&	6&	6	\\
\#19 &			RICAWEBSERVICES &	26832 &		733K (0.11\%)&4&	4&	2	\\
\#20 &			ORACLE-BMC &		31898 &		676K (0.10\%)&42&	38&	34	\\
			\bottomrule
	\end{tabular}
\end{table}

\received{December 2024}
\received[revised]{June 2025}
\received[accepted]{June 2025}

\end{document}